%% file: PAPER.tex
\documentclass[10pt,twocolumn,twoside]{IEEEtran}

\usepackage{amsmath}
\usepackage{mathtools}
\usepackage{tikz}
\usepackage{verbatim}
\usepackage{graphicx}
\usepackage{epstopdf}
\usepackage{amssymb}
\usepackage{relsize}
\usepackage[textwidth=3.8em,textsize=scriptsize]{todonotes}
\usepackage{algorithm}
\definecolor{Gray}{gray}{0.90}
\usepackage{colortbl}
\usepackage{algorithmicx}
\newtheorem{theorem}{\bf{Theorem}}[section]

\newtheorem{prop}[theorem]{\bf{Proposition}}

\newenvironment{definition}[1][Definition]{\begin{trivlist}
\item[\hskip \labelsep {\bfseries #1}]}{\end{trivlist}}
\def\qed{\hfill\rule[-1pt]{5pt}{5pt}\par\medskip}
\usepackage{amssymb}
\usepackage{url}
\DeclareMathOperator*{\argmax}{arg\!\max}
\DeclareRobustCommand{\c4c4}{
\scalebox{0.3}{
\begin{tikzpicture}
\draw (0,0) -- (0.5,0.5); \draw (0.5,0.5) -- (1,0); \draw (1,0) -- (0.5,-0.5);
\draw (0.5,-0.5) -- (0,0); \draw (1,0) -- (1.5,0.5); \draw (1.5,0.5) -- (2,0);
\draw (2,0) -- (1.5,-0.5); \draw (1.5,-0.5) -- (1,0); \filldraw[black] (0,0) circle (3pt);
\filldraw[black] (0.5,0.5) circle (3pt); \filldraw[black] (1,0) circle (3pt);
\filldraw[black] (0.5,-0.5) circle (3pt); \filldraw[black] (2,0) circle (3pt);
\filldraw[black] (1.5,0.5) circle (3pt); \filldraw[black] (1.5,-0.5) circle (3pt);
\end{tikzpicture}%
}}
\DeclareRobustCommand{\cycle4}{
\scalebox{0.3}{
\begin{tikzpicture}
\draw (0,0) -- (0.5,0.5); \draw (0.5,0.5) -- (1,0); \draw (1,0) -- (0.5,-0.5);
\draw (0.5,-0.5) -- (0,0); 
\filldraw[black] (0,0) circle (3pt);\filldraw[black] (0.5,0.5) circle (3pt); 
\filldraw[black] (1,0) circle (3pt); \filldraw[black] (0.5,-0.5) circle (3pt);
\end{tikzpicture}%
}}
\DeclareRobustCommand{\star}{
\scalebox{0.3}{
\begin{tikzpicture}
\draw (0,0) -- (0.8,0); \draw (0,0) -- (-0.8,0); 
\draw (0,0) -- (0.4,0.708); \draw (0,0) -- (-0.4,0.708);
\draw (0,0) -- (-0.4,-0.708); \draw (0,0) -- (0.4,-0.708);
\filldraw[black] (0,0) circle (3pt);
\filldraw[black] (0.8,0) circle (3pt); \filldraw[black] (-0.8,0) circle (3pt);
\filldraw[black] (0.4,0.708) circle (3pt); \filldraw[black] (-0.4,0.708) circle (3pt);
\filldraw[black] (-0.4,-0.708) circle (3pt); \filldraw[black] (0.4,-0.708) circle (3pt);
\end{tikzpicture}%
}}

\DeclareRobustCommand{\cyc}{
\scalebox{0.3}{
\begin{tikzpicture}
\draw (0,0) -- (0.5,0.5); \draw (0.5,0.5) -- (1,0.5); \draw (1,0.5) -- (1.5,0.5);
\draw (1.5,0.5) -- (1.5,-0.5);  \draw (1.5,-0.5) -- (1,-0.5); \draw (1,-0.5) -- (0.5,-0.5);
\draw (0.5,-0.5) -- (0,0);
\filldraw[black] (0,0) circle (3pt);\filldraw[black] (0.5,0.5) circle (3pt); 
\filldraw[black] (1,0.5) circle (3pt); \filldraw[black] (1.5,0.5) circle (3pt);
\filldraw[black] (1.5,-0.5) circle (3pt); \filldraw[black] (1,-0.5) circle (3pt);
\filldraw[black] (0.5,-0.5) circle (3pt);
\end{tikzpicture}%
}}

\DeclareRobustCommand{\A}{
\scalebox{0.3}{
\begin{tikzpicture}
\draw (0,0) -- (0.5,0.5); \draw (0.5,0.5) -- (1,0.5); \draw (1,0.5) -- (1.5,0.5);
\draw (1.5,0.5) -- (1.5,-0.5);  \draw (1.5,-0.5) -- (1,-0.5); \draw (1,-0.5) -- (0.5,-0.5);
\draw (0.5,-0.5) -- (0,0); \draw (1,0.5) -- (1,-0.5);
\filldraw[black] (0,0) circle (3pt);\filldraw[black] (0.5,0.5) circle (3pt); 
\filldraw[black] (1,0.5) circle (3pt); \filldraw[black] (1.5,0.5) circle (3pt);
\filldraw[black] (1.5,-0.5) circle (3pt); \filldraw[black] (1,-0.5) circle (3pt);
\filldraw[black] (0.5,-0.5) circle (3pt);
\end{tikzpicture}%
}}
\DeclareRobustCommand{\B}{
\scalebox{0.3}{
\begin{tikzpicture}
\draw (0,0) -- (0.5,0.5); \draw (0.5,0.5) -- (1,0.5); \draw (1,0.5) -- (1.5,0.5);
\draw (1.5,0.5) -- (1.5,-0.5);  \draw (1.5,-0.5) -- (1,-0.5); \draw (1,-0.5) -- (0.5,-0.5);
\draw (0.5,-0.5) -- (0,0); \draw (0.5,0.5) -- (1,-0.5); \draw (1,0.5) -- (0.5,-0.5);
\filldraw[black] (0,0) circle (3pt);\filldraw[black] (0.5,0.5) circle (3pt); 
\filldraw[black] (1,0.5) circle (3pt); \filldraw[black] (1.5,0.5) circle (3pt);
\filldraw[black] (1.5,-0.5) circle (3pt); \filldraw[black] (1,-0.5) circle (3pt);
\filldraw[black] (0.5,-0.5) circle (3pt);
\end{tikzpicture}%
}}
\DeclareRobustCommand{\C}{
\scalebox{0.3}{
\begin{tikzpicture}
\draw (0,0) -- (0.5,0.5); \draw (0.5,0.5) -- (1,0); \draw (1,0) -- (0.5,-0.5);
\draw (0.5,-0.5) -- (0,0); \draw (1,0) -- (1.5,0.5); \draw (1.5,0.5) -- (2,0);
\draw (2,0) -- (1.5,-0.5); \draw (1.5,-0.5) -- (1,0); \draw (0.5,-0.5) -- (1.5,-0.5);
\filldraw[black] (0,0) circle (3pt);
\filldraw[black] (0.5,0.5) circle (3pt); \filldraw[black] (1,0) circle (3pt);
\filldraw[black] (0.5,-0.5) circle (3pt); \filldraw[black] (2,0) circle (3pt);
\filldraw[black] (1.5,0.5) circle (3pt); \filldraw[black] (1.5,-0.5) circle (3pt);
\end{tikzpicture}%
}}

\DeclareRobustCommand{\D}{
\scalebox{0.3}{
\begin{tikzpicture}
\draw (0,0) -- (0.5,0.5); \draw (0.5,0.5) -- (1,0.5); \draw (1,0.5) -- (1.5,0.5);
\draw (1.5,0.5) -- (1.5,-0.5);  \draw (1.5,-0.5) -- (1,-0.5); \draw (1,-0.5) -- (0.5,-0.5);
\draw (0.5,-0.5) -- (0,0); \draw (0.5,0.5) -- (1,-0.5); \draw (1,0.5) -- (0.5,-0.5);
\draw (1,0.5) -- (1,-0.5);
\filldraw[black] (0,0) circle (3pt);\filldraw[black] (0.5,0.5) circle (3pt); 
\filldraw[black] (1,0.5) circle (3pt); \filldraw[black] (1.5,0.5) circle (3pt);
\filldraw[black] (1.5,-0.5) circle (3pt); \filldraw[black] (1,-0.5) circle (3pt);
\filldraw[black] (0.5,-0.5) circle (3pt);
\end{tikzpicture}%
}}
\DeclareRobustCommand{\K}{
\scalebox{0.3}{
\begin{tikzpicture}
\draw (0,0.25) -- (1,0); \draw (0,0.25) -- (1,0.5); \draw (0,0.25) -- (1,-0.5); 
\draw (0,-0.25) -- (1,0); \draw (0,-0.25) -- (1,0.5); \draw (0,-0.25) -- (1,-0.5); 
\filldraw[black] (0,0.25) circle (3pt);\filldraw[black] (0,-0.25) circle (3pt);
\filldraw[black] (1,0) circle (3pt);\filldraw[black] (1,0.5) circle (3pt); 
\filldraw[black] (1,-0.5) circle (3pt); 
\end{tikzpicture}%
}}

\begin{document}

\title{Scheduling Resource-Bounded Monitoring Devices for Event Detection and Isolation in Networks} 
\author{Waseem~Abbas, 
        Aron~Laszka, 
        Yevgeniy~Vorobeychik, and 
        Xenofon~Koutsoukos
\thanks{W.~Abbas is with the Institute for Software Integrated Systems, Vanderbilt University, Nashville, TN, 37212 USA (email: waseem.abbas@vanderbilt.edu)}
\thanks{A.~Laszka is with the Department of Electrical Engineering and Computer Science at the University of California, Berkeley, CA 94720, USA (email: laszka@berkeley.edu).}
\thanks{Y.~Vorobeychik, and X.~Koutsoukos are with the Department of Electrical Engineering and Computer Science, Vanderbilt University, and also with the Institute for Software Integrated Systems, Vanderbilt University, Nashville, TN 37212 (emails: yevgeniy.vorobeychik@vanderbilt.edu, xenofon.koutsoukos@vanderbilt.edu).}}
\markboth{Journal of \LaTeX\ Class Files,~Vol.~XX, No.~X, June~2016}%
{Abbas \MakeLowercase{\textit{et al.}}: Scheduling Resource Bounded Monitoring Devices}
\maketitle
\begin{abstract}
In networked systems, monitoring devices such as sensors are typically deployed to monitor various target locations. Targets are the points in the physical space at which events of some interest, such as random faults or attacks, can occur. Most often, these devices have limited energy supplies, and they can operate for a limited duration. As a result, energy-efficient monitoring of various target locations through a set of monitoring devices with limited energy supplies is a crucial problem in networked systems.
In this paper, we study optimal scheduling of monitoring devices to maximize network coverage for detecting and isolating events on targets for a given network lifetime. 
The monitoring devices considered could remain active only for a fraction of the overall network lifetime. 
We formulate the problem of scheduling of monitoring devices as a graph labeling problem, which unlike other existing solutions, allows us to directly utilize the underlying network structure to explore the trade-off between coverage and network lifetime. In this direction, first we propose a greedy heuristic to solve the graph labeling problem, and then provide a game-theoretic solution to achieve optimal graph labeling. 
Moreover, the proposed setup can be used to simultaneously solve the scheduling and placement of monitoring devices, which yields improved performance as compared to separately solving the placement and scheduling problems. 
Finally, we illustrate our results on various networks, including real-world water distribution networks. 
\end{abstract}
\begin{IEEEkeywords}
scheduling, networked systems, network coverage, graph labeling, potential games, dominating sets.
\end{IEEEkeywords}
\section{Introduction}
\input{intro}

\section{System Model and Problem Formulation}
\label{sec:model}
In this section, first, we present the system model by describing all the major components involved, and then we formulate the problem of optimal scheduling of resource bounded monitoring devices in networks.

\textit{(a) Network Graph --} We model the network as an \textit{undirected graph}, $G(V,E)$, in which $V$ is the set of nodes, and $E$ is the set of edges given by the unordered pairs of nodes. Two nodes are adjacent if there exists an edge between them. The \textit{neighborhood of a node} $v$, denoted by $N(v)$, is the set of all nodes that are adjacent to $v$, i.e., $N(v) = \{u: \; (u,v) \in E\}$, and the \textit{neighborhood of a subset of nodes $S$}, denoted by $N(S)$, is $\bigcup\limits_{v\in S} N(v)$. 
The \textit{degree} of a node $v$, represented by $\delta(v)$, is simply $\delta(v) = |N(v)|$. Moreover, a \textit{path} is a sequence of nodes such that any two consecutive nodes in the path are adjacent, and the number of edges included in the path is the \textit{length} of the path. 
Any two nodes are said to be \textit{connected} if there exists a path between them. The \textit{distance between connected nodes} $u$ and $v$, denoted by $d(u,v)$, is the length of the shortest path between them. Similarly, the \textit{distance between node $u$ and edge $e = (i,j)$} is $d(u,e) = \max(d(u,i),d(u,j))$. The network graph abstracts interactions among various nodes within the network. 

\textit{(b) Targets --} They are a subset of nodes and/or edges, denoted by $Y\subseteq (V\cup E)$, that could be subjected to an abnormal activity (or \textit{event}), such as pipe failure, and therefore, need to be monitored by monitoring devices. 

{\textit{(c) Monitoring Devices --}} These are the devices that are deployed at a subset of nodes $S\subseteq V$ in the network, and can monitor the other nodes and/or links within the network for any unusual activity, for instance, link failure detection such as pipe burst in water networks. We refer to any such abnormal activity on a target as an \textit{event}. A monitoring device can monitor all nodes and edges for events that lie within some pre-specified distance, referred to as the \textit{range}, of the device. If $u$ is the node at which a monitoring device with the range $\lambda$ is deployed, then the device \textit{covers} (monitors) all the nodes and edges in the set 
$$
\{v\subseteq V:\;d(u,v)\le \lambda\} \cup \{e\subseteq E:\;d(u,e)\le \lambda\}.
$$

In other words, a target is \textit{covered} if and only if it lies within the range of some monitoring device. Each device is resource-bounded in terms of the available \textit{battery supply}, denoted by $B$, which means that a device can be \textit{active} (or can be operational) for only $B$ time duration. Furthermore, a monitoring device has only \textit{two output} states -- event detected at some target without knowing the exact location of the target, and no event detected.

\subsection{Network Performance Measures}\label{sec:measures}
We are interested in measuring the quality of monitoring of targets through a set of monitoring devices, both from the detection and isolation perspectives. In \textit{detection}, the objective is just to detect any abnormal activity on some target irrespective of determining the exact location of it, whereas in \textit{isolation}, the goal is to \textit{uniquely detect} the target at which the abnormal activity occurs. Moreover, we refer to the overall lifetime of the network, i.e., duration for which monitoring of targets for detection (isolation) is considered, as the \textit{network lifetime} $T$. To simplify, we divide the time into \textit{time slots} of equal length. The battery supply $B$ of a monitoring device could be represented by the number of time slots, say $\sigma$, in which the device could remain active. 
Moreover, the network lifetime $T$ could be represented by the total number of time slots, say $k$, for which the detection (isolation) of targets is considered. 
Note that $T$ and $B$ represent the actual \textit{duration} of overall network lifetime and battery lifetime of individual monitoring device respectively, whereas, $k$ and $\sigma$, which are chosen to be positive integers, represent respectively the total \textit{number of time slots} and the time slots for which each device could remain active. 


\textit{(a) Detection Measure -- }Let there be a total of $m$ targets, and $m_i$ be the number of targets that are covered by the monitoring devices that are active in the $i^{th}$ time slot. We define the \textit{average detection performance}, denoted by $\mathcal{D}$, as 
\begin{equation}
\label{eq:P}
\mathcal{D} = \frac{1}{k}\sum\limits_{i=1}^{k} \left(\frac{m_i}{m}\right).
\end{equation}

\textit{(b) Isolation Measure -- }We observe that event at target $i$ can be distinguished from an event at target $j$ if and only if there exists a monitoring device that gives different outputs in case of events at $i$ and $j$. In other words, there exists a monitoring device at some node such that exactly one of the target (either $i$ or $j$, but not both) is covered by the monitoring device. If such a monitoring device exists, we say that the \textit{target-pair $i,j$ is covered}. The event at target $i$ can be uniquely detected (or can be distinguished from events at all other targets) if all target-pairs $i,j$ ($j\ne i$) are covered. If $m$ is the total number of targets, then there is a total of $\ell = \dbinom{m}{2}$ target-pairs. In the $j^{th}$ time slot, let $\ell_j$ be the number of target-pairs that are covered by the active sensors. Then, we define the \textit{average isolation performance}, denoted by $\mathcal{I}$, as 

\begin{equation}
\label{eq:I}
\mathcal{I} = \frac{1}{k}\sum\limits_{j=1}^{k} \left(\frac{\ell_j}{\ell}\right)
\end{equation}
\noindent
where $k$ is the total number of time slots. A list of symbols used throughout the paper is given in Table \ref{tab:symbols}.

\begin{table}
\caption{List of Symbols}
\label{tab:symbols}
\centering
\renewcommand*{\arraystretch}{1.5}
\begin{tabular}{| l | l |}
\hline
Symbol & Description \\
\hline\hline
$G(V,E)$ & network graph \\
\rowcolor{Gray} $S$ & set of monitoring devices ($S\subseteq V$) \\
$Y$ & set of targets $(Y\subseteq (V\cup E))$\\
\rowcolor{Gray} $\lambda$ & range of monitoring device \\
$N(v)$ & neighborhood of a node $v$ \\
\rowcolor{Gray} $N(S)$ & neighborhood of a subset of nodes $S$ \\
$T$ & network lifetime in terms of actual time duration \\
\rowcolor{Gray} $B$ & duration for which a device can remain active  \\
$k$ & network lifetime in terms of the total number of time slots \\
\rowcolor{Gray} $\sigma$ & number of time slots in which a device can remain active \\
$\mathcal{D}$ & average detection measure (\ref{eq:P}) \\
\rowcolor{Gray} $\mathcal{I}$ & average isolation measure (\ref{eq:I}) \\
$S_i$ & nodes at which devices are active in the $i^{th}$ time slot\\
\rowcolor{Gray} $\mathcal{G}(\mathcal{V},\mathcal{X})$ & bi-partite graph representation of the network $G(V,E)$\\
\hline
\end{tabular}
\end{table}

\subsection{Problem Formulation}
\label{sec:Problem}
Consider a network $G(V,E)$ in which $S\subseteq V$ is the subset of nodes at which monitoring devices with ranges $\lambda$ are deployed, and $Y\subseteq (V\cup E)$ are the set of targets. Each monitoring device could remain active in at most $\sigma$ of the total of $k$ time slots due to battery supply constraints. In each time slot $i$, let $S_i\subseteq S$ be the subset of nodes with active monitoring devices. Thus, we get a \textit{schedule} of (active) monitoring devices as $S_1,S_2,\cdots,S_k$.

\textit{The objective is to determine the maximum average detection performance $\mathcal{D}$ (or average isolation performance $\mathcal{I}$) for   a given network life time, represented by $k$ time slots, under the battery constraints of monitoring devices, represented by $\sigma$ time slots, and also a schedule of monitoring devices that achieves the maximum $\mathcal{D}$ (or $\mathcal{I}$).}

It is obvious that as $k$ increases, the maximum values of $\mathcal{D}$ (or $\mathcal{I}$) decrease. So, in a way, our goal is to understand a relationship between $k$ and $\mathcal{D}$ (or $\mathcal{I}$), and design a systematic scheme to obtain a schedule for activating monitoring devices with limited battery supplies to obtain the desired network performance. Note that the scheduling problem for a \textit{complete coverage} of targets, in which the objective is to determine a schedule that ensures $\mathcal{D}=1$ throughout the network life is a special case of the above problem.


\input{Complexity}

\input{Labeling}
\input{Game_Theoretic}

\input{New_Simultaneous_Placement}
\input{Simulations}
\input{Special_Cases}

\input{Related_Work}

\section{Conclusions}
\label{sec:con}
We studied the problem of scheduling resource bounded monitoring devices in networks to maximize the  detection and isolation of failure events for a given network lifetime. We showed that the scheduling problem is equivalent to a graph labeling problem, which allowed direct exploitation of the network structure to obtain optimal schedules. To solve the graph labeling problem, we presented a game-theoretic solution. 
We also showed that the detection (isolation) performance of monitoring devices deployed within the network was better when the placement and scheduling problems for these devices were solved simultaneously as compared to the case in which the optimal placement of these devices was solved first followed by the computation of optimal schedules. Our graph labeling formulation and game-theoretic solution allowed us to simultaneously solve placement and scheduling problems. We demonstrated results for various networks including water distribution and random networks. The graph labeling problem presented here could be useful in solving resource allocation problems in other domains such as multi-agent and multi-robot systems. Moreover, the proposed approach could be effective in characterizing and comparing network topologies in terms of the coverage performance,  
i.e., using resource-constraint monitoring devices, which network structures could result in higher detection and isolation performances? 

\appendix[Proof of Theorem \ref{thm:random}]
\label{app}
The average detection performance, in this case, is equivalent to finding the probability that an arbitrary node $u$ is covered in an arbitrary time slot $i$. In this direction, we observe that 

\begin{equation}
\label{pf_1}
\begin{split}
\text{Pr } & \left(u \text{ is not covered in the } i^{th} \text{ slot}\right) \\
& = \text{Pr} \left(\begin{array}{cc}u \text{ is not active}\\ \text{in the } i^{th} \text{ slot}\end{array}\right) \prod_{v\in N(u)} \text{Pr } \left(\begin{array}{cc}v \text{ is not active}\\ \text{in the } i^{th} \text{ slot}\end{array}\right)\\
\end{split}
\end{equation}
Here, 
\begin{equation}
\label{pf_0}
\begin{split}
\text{Pr} \left(\begin{array}{cc}u \text{ is not active}\\ \text{in the } i^{th} \text{ slot}\end{array}\right)  = 1 - \frac{\dbinom{\sigma-1}{k-1}}{\dbinom{k}{\sigma}} = \frac{k-\sigma}{k}.
\end{split}
\end{equation}

The second term in (\ref{pf_1}) is the probability that none of the nodes in the neighborhood of node $u$ is active in the $i^{th}$ time slot. The probability of having $j$ neighbors in $N(u)$ in a random geometric graph is given by Poisson distribution, i.e., $\frac{\left(\lambda \pi r^2\right)^j e^{-\lambda \pi r^2}}{j!}$. Thus,

\begin{equation}
\label{pf_2}
\begin{split}
\prod_{v\in N(u)} & \text{Pr } \left(v \text{ is not active in the } i^{th} \text{ slot}\right)\\
& = \sum\limits_{j=0}^{\infty} \frac{\left(\lambda \pi r^2\right)^j e^{-\lambda \pi r^2}}{j!} \left(\frac{k-\sigma}{k}\right)^j
\end{split}
\end{equation}

Inserting (\ref{pf_0}) and (\ref{pf_2}) in (\ref{pf_1}), we get

\begin{equation}
\label{pf_3}
\begin{split}
\text{Pr } & \left(u \text{ is not covered in the } i^{th} \text{ slot}\right) \\
& = e^{-\lambda \pi r^2}\frac{(k-\sigma)}{k}
\sum\limits_{j=0}^{\infty} \frac{1}{j!} \left(\frac{\lambda \pi r^2(k-\sigma)}{k}\right)^j\\
& = e^{-\lambda \pi r^2}\frac{(k-\sigma)}{k} e^{\frac{\lambda \pi r^2(k-\sigma)}{k}}\\
& = \left(\frac{k-\sigma}{k}\right)e^{\frac{-\sigma\lambda\pi r^2}{k}}
\end{split} 
\end{equation}.

Thus, we get the desired result as

\begin{equation*}
\begin{split}
\text{Pr } \left(u \text{ is covered in the } i^{th} \text{ slot}\right) = 1 - \left(\frac{k-\sigma}{k}\right)e^{\frac{-\sigma\lambda\pi r^2}{k}}
\end{split}
\end{equation*}
\qed

\bibliographystyle{IEEEtran}
\bibliography{references}

\end{document}

%% file: intro.tex
Detection and isolation of unwanted events such as faults, failures, and malicious intrusions is a fundamental concern in a variety of practical networks. For example, leakage detection in water distribution networks can reduce physical damage as well as financial losses \cite{farley2003losses}. For this purpose, monitoring devices, such as sensors, are typically deployed strategically throughout the network. 
Spatially distributed systems over large areas may often be monitored only by battery-powered devices, as wired deployment can be prohibitively expensive or impossible. If the power supply provided by batteries is insufficient for continuous monitoring during the intended lifetime of a system, batteries must be replaced regularly.
Since the cost of battery replacement for a large number of devices can be very expensive, one of the primary design concerns for such systems is increasing the time until the batteries of the sensors are depleted. At the same time, it is desired to maintain a certain level of monitoring in terms of the number of targets covered throughout the network lifetime. Here, targets are the points in the physical space at which events of interest can occur. For instance, in water distribution networks, events can be the pipe bursts, and so targets can be the water pipes, which need to be monitored through sensors such as battery operated pressure sensors.

One of the primary approaches for conserving battery power is ``sleep scheduling.''
The idea is to have only a subset of the sensors activated at any given time, and to turn off (i.e., ``sleep'') the remaining ones, thereby conserving power.
By activating different sets of devices one after another, the overall lifetime of a system can be substantially increased. 
Previous works have mostly focused on finding schedules that ensure complete coverage, that is, guaranteeing that every target is monitored by some device at any given moment in time (e.g., \cite{cardei2005energy,Wang:2011:CPS:1978802.1978811}).
However, complete coverage is a very strict requirement, which severely limits the sets of devices that may be asleep at the same time.
In fact, coverage (i.e., ratio of monitored targets to the total number of targets) is a submodular function of the set of active devices in most models (e.g., \cite{perelman2015sensor,krause2008efficient}), which roughly means that attaining complete coverage is disproportionately expensive as compared to achieving reasonably good coverage. Managing energy resources of monitoring devices via their scheduling to achieve an appropriate coverage of targets is a significant issue in networks where extended network lifetime is a critical requirement.

In this paper, we design efficient scheduling schemes for a set of monitoring devices with limited battery supplies to achieve maximum target coverage for a given network lifetime. 
Scheduling of such devices to achieve complete network coverage is a special case of this general formulation. We model the network as a graph, in which monitoring devices could be deployed at a subset of nodes, and the targets could be nodes and/or edges. Each monitoring device has a limited active time, and covers a subset of targets within its range during its active time. For a given network lifetime, the objective is to determine the maximum possible coverage, both in terms of the detection and isolation of (events at) targets, and a schedule of monitoring devices to obtain the optimal coverage.

In this direction the main contributions of the paper are:

\noindent
(1) We show that the optimal scheduling of monitoring devices is an APX-hard problem, that is, there is no polynomial-time approximation scheme (PTAS) for the problem unless P=NP. 

\noindent
(2) We provide a graph-theoretic formulation of the scheduling problem by showing that it is equivalent to a unique graph labeling problem, which allows us to directly exploit the network structure to obtain optimal schedules. 

\noindent
(3) To solve the graph labeling, and hence the scheduling problem, we propose two solutions; first, a greedy heuristic that runs in polynomial time, and gives near optimal solutions for many networks as we illustrate. However, in general, performance guarantees of the heuristic in terms of the optimality of the solution remain unknown. Second, we present a game-theoretic solution, in which we show that the labeling problem can be posed as a potential game, for which efficient algorithms, such as binary log-linear learning (BLLL), are known that asymptotically give globally optimal solutions with an arbitrary high probability. 

\noindent
(4) Moreover, we illustrate that the game-theoretic solution allows simultaneously optimizing the placement and scheduling of monitoring devices, which gives better results as compared to separately solving the placement and scheduling. Note that the placement problem involves selecting optimal locations to deploy a given set of monitoring devices to maximize the target coverage within networks. 

\noindent
(5) We analyze the performance of the approach through simulations on various networks including real-world water distribution networks and random networks. For random networks, we also provide analytical results to determine the performance of random scheduling, which does not require any information about the network structure. 

\noindent
(6) Finally, we consider some practically relevant special cases of the problem, such as scheduling to maximize network lifetime while ensuring complete coverage of the targets within the network.

The rest of the paper is organized as follows: In Section \ref{sec:model}, we explain our system model and define the scheduling problem. Section \ref{sec:ref} addresses the issue of complexity of the problem. In Section \ref{sec:labeling}, we present a graph labeling based formulation of the scheduling, and in Section \ref{sec:game_sol} propose solutions to the graph labeling problem. Section \ref{sec:simulations} illustrates simulation results, and section \ref{sec:special} presents a particular case of interest of the scheduling problem. In Section \ref{sec:related_work}, we provide an overview of related work, and conclude the paper in Section \ref{sec:con}.

%% file: Complexity.tex
\section{Problem Complexity}
\label{sec:ref}
In this section, we show that the problem of finding a schedule that maximizes the average detection performance for a given network lifetime and battery supplies, as discussed in Section \ref{sec:Problem}, is APX-hard. APX-hardness implies that (unless P=NP), there does not exist a polynomial-time algorithm that can solve the problem to within arbitrary multiplicative factor of the optimum. 


In our case, for a target $\tau$, if $Q_\tau$ represents the fraction of the total number of time slots in which an event on $\tau$ can be detected (i.e., $\tau$ is covered), then the expected value of detecting an event on an arbitrary target, denoted by $\mathcal{Q}$ is
\begin{equation}
\label{eq:Q}
\mathcal{Q} = \frac{1}{|Y|} \sum_{\tau\in Y} Q_\tau .
\end{equation}

Note that $\mathcal{Q}$ and $\mathcal{D}$ have exactly same values for a given schedule $(S_1,S_2,\cdots,S_k)$, and therefore, they both measure the average detection performance of the schedule. 
We formulate finding a schedule that maximizes detection performance as the following optimization problem:

\begin{definition}\textit{(Maximum Average Detection):}
Given a graph $G = (V, E)$, a set of monitoring devices $S \subseteq V$, a set of targets $Y\subseteq (V\cup E)$, range of the monitoring device $\lambda$, a network lifetime represented by $k$ time slots, a battery supply represented by $\sigma$ time slots, find a schedule $(S_1, S_2, \ldots, S_k)$ that maximizes the average detection performance $\mathcal{Q}$.
\end{definition}

\begin{theorem}
\label{thm:complexity}
The Maximum Average Detection Problem is APX-hard.
\end{theorem}

We show APX-hardness by reducing a well-known APX-hard problem, the Maximum Cut Problem \cite{papadimitriou1988optimization} to the detection problem.
The Maximum Cut Problem is defined as follows:

\begin{definition}\textit{(Maximum Cut Problem):}
Given a graph $G= (V, E)$, find a disjoint partition $V_1, V_2$ of $V$ that maximizes the number of edges $|E(V_1, V_2)|$ between $V_1$ and $V_2$.
\end{definition}

\textit{Proof (Theorem~\ref{thm:complexity}) -- }
We prove APX-hardness by showing that there exists a PTAS-reduction from the Maximum Cut Problem to the Maximum Average Detection Problem. 
First, we define a polynomial-time mapping from an instance of the cutting problem to an instance of the detection problem:
\begin{itemize}
\item let the network of the Maximum Average Detection Problem be the graph of the Maximum Cut Problem;
\item let the set of monitoring devices be $S = V$;
\item let the set of targets be $Y = E$;
\item let the range of the monitoring device be $\lambda = 1$;
\item let the network lifetime be $k = 2$ time slots;
\item and let the battery supply be $\sigma = 1$ time slot.
\end{itemize}

Second, we define a polynomial-time mapping from a solution $(S_1, S_2)$ of an instance of the detection problem (i.e., a schedule) to a solution $(V_1, V_2)$ of the corresponding instance of the cutting problem (i.e., a cut):
\begin{equation}
V_1 := S_1 \text{ and } V_2 := S_2 .
\end{equation}
Next, observe that if an edge is cut by $(V_1, V_2)$, then the corresponding target is covered by both $S_1$ and $S_2$, which implies $Q_\tau = 1$.
On the other hand, if an edge is not cut by $(V_1, V_2)$, then the corresponding target is covered in only one time slot, which implies $Q_\tau = \frac{1}{2}$.
Consequently, for any pair of solutions $(S_1, S_2)$ and $(V_1, V_2)$, we have
\begin{align}
\mathcal{Q}(S_1, S_2) &= \frac{1}{|E|} \left( \sum_{\tau \in E(V_1, V_2)} 1 + \sum_{\tau \not\in E(V_1, V_2)} \frac{1}{2} \right) \\
 &= \frac{1}{2} + \frac{1}{2}\frac{|E(V_1, V_2)|}{|E|} .
\end{align}

Using the same argument, we can also show that if a schedule $(S_1, S_2)$ is an optimal solution to the detection problem, then the cut $(V_1 = S_1, V_2 = S_2)$ is also an optimal solution to the cutting problem, and vice versa.
Therefore, if a schedule $(S_1, S_2)$ is at most $(1 - \epsilon)$ times worse than the optimal schedule, then the corresponding cut $(V_1, V_2)$ is at most $(1 - 2\epsilon)$ times worse than the optimal cut.
Consequently, there is a PTAS-reduction from the Maximum Cut Problem to the Maximum Average Detection Problem. 
\qed

As a consequence, we cannot optimally solve the maximum average detection problem in a polynomial time. Hence, we need efficient heuristics that can provide reasonably good solutions with acceptable time complexities. In this regard, it becomes crucial to maximally exploit the structure of the problem in a systematic way. To achieve this objective, we first provide a graph-theoretic formulation of the scheduling problem in the next section, and then provide efficient solution to the problem using a game-theoretic setting in Section \ref{sec:game_sol}.

%% file: Labeling.tex
\section{A Graph-Theoretic Formulation of the Scheduling Problem}
\label{sec:labeling}
In this section, using various graph-theoretic notions, we formulate the scheduling problem as a graph labeling problem. In the next section, a solution approach is presented to solve the corresponding graph labeling, thus solving the the original scheduling problem.

Our approach is to first obtain a \textit{bi-partite graph}, denoted by $\mathcal{G}(\mathcal{V},\mathcal{E})$, from a given graph. This bi-partite graph illustrates targets and the monitoring devices with given ranges covering those targets. We then formulate the scheduling problem on the original network $G(V,E)$ as a graph labeling problem on the bi-partite graph $\mathcal{G}(\mathcal{V},\mathcal{E})$.

\subsection{Bi-partite Graphs in the Cases of Detection and Isolation}
\label{sec:labeling_A} 
\subsubsection{Case 1 -- Detection}
When scheduling of monitoring devices is required with an objective to maximize the average detection score $\mathcal{D}$, as described in Section \ref{sec:measures}, the bi-partite graph $\mathcal{G}(\mathcal{V},\mathcal{E})$ is simply obtained as follows: the vertex set $\mathcal{V}$ is the union $\mathcal{X}\cup\mathcal{Y}$, where $\mathcal{X}=S\subseteq V$ is the set of nodes corresponding to the set of monitoring devices, and $\mathcal{Y} = Y$ is the set of targets in the original network $G$. Moreover, each $x\in\mathcal{X}$ is adjacent to vertices in $\mathcal{Y}$ that are at most $\lambda$ distance away from $x$ in $G$. An example is shown in Figure \ref{fig:example}.

\subsubsection{Case 2 -- Isolation}
If maximizing the average isolation measure $\mathcal{I}$, as in Section \ref{sec:measures}, is the objective of scheduling, then $\mathcal{G}(\mathcal{V},\mathcal{E})$ is obtained as follows: As in the case of detection, the vertex set of the bi-partite graph is $\mathcal{V} = \mathcal{X}\cup\mathcal{Y}$, where $\mathcal{X}=S\subseteq V$ corresponds to the set of monitoring devices. To obtain $\mathcal{Y}$, we define a node for every pair of targets in $Y$. There will be $\dbinom{|Y|}{2}$ such nodes in $\mathcal{Y}$. As for the edge set $\mathcal{E}$ of the bi-partite graph, let $y\in \mathcal{Y}$ corresponds to the (unordered) target pair $(\tau_1,\tau_2)\in Y$. 
Then, each $x\in\mathcal{X}$ is adjacent to $y\in\mathcal{Y}$ in $\mathcal{G}$ if and only if exactly one of the targets $\tau_1$ or $\tau_2$ is within $\lambda$ distance from (the monitoring device corresponding to) $x$ in the original network $G$. In other words, in the bi-partite graph $\mathcal{G}$, there will be no edge between $x$ and $y$ that corresponds to the target pair $(\tau_1,\tau_2)$, if and only if the monitoring device $x$ covers both targets $\tau_1$ and $\tau_2$ in $G$, or does not cover any of the targets $\tau_1$ and $\tau_2$. An example is illustrated in Figure \ref{fig:example}.

\subsubsection*{Example}
\begin{figure}[htb]\begin{center}
\includegraphics[scale=0.6]{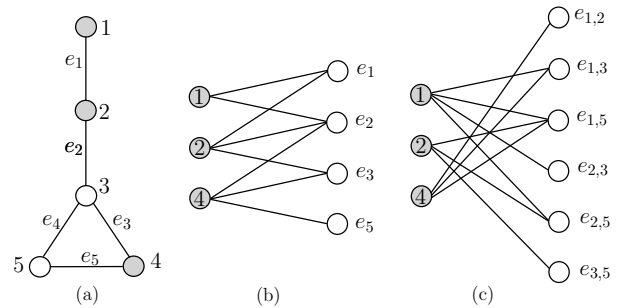}
\caption{(a) An example network graph $G(V,E)$. Bi-partite graph representations for (b) detection and (c) isolation.} 
\label{fig:example}\end{center}\end{figure}

Consider a graph $G(V,E)$ in Figure \ref{fig:example}. Let $S = \{1,2,4\}\subseteq V$ be the set of monitoring devices and edges in the set $Y = \{e_1,e_2,e_3,e_5\}$ be the targets. Moreover, each monitoring device has the range $\lambda = 2$. The bi-partite graphs $\mathcal{G}(\mathcal{V},\mathcal{E})$ for the scheduling of monitoring devices to maximize the detection and isolation measures are shown in Figures \ref{fig:example}(b) and \ref{fig:example}(c) respectively. The vertex set of bi-partite graphs in both cases is $\mathcal{V} = \mathcal{X}\cup \mathcal{Y}$, where $\mathcal{X} = S$. For the detection case, $\mathcal{Y} = Y$, and for the isolation case, $\mathcal{Y} = \{e_{12},e_{13},e_{15},e_{23},e_{25},e_{35}\}$, where $e_{ij}$ corresponds to the pair of edges $(e_i,e_j)$ in $Y$. Note that an edge between $x\in\mathcal{X}$ and $e_{ij}\in\mathcal{Y}$ indicates that the monitoring device at $x$ covers the target pair $(e_i,e_j)$, or in other words, can distinguish between events at $e_i$ and $e_j$.

\subsection{A Graph Labeling Problem and its Equivalence to the Scheduling Problem}
After obtaining the bi-partite graph $\mathcal{G}(\mathcal{V}=\mathcal{X}\cup\mathcal{Y},\mathcal{E})$ from a given network $G(V,E)$, we can re-write the detection and isolation scores as in (\ref{eq:P}) and (\ref{eq:I}) respectively in terms of $\mathcal{G}$. Note that if $S_i\subseteq  \mathcal{X}$ is the subset of active monitoring devices in the $i^{th}$ time slot, then for the detection (isolation), the set of targets (target-pairs) covered by $S_i$ is simply the neighborhood of set $S_i$, i.e., $N(S_i) = \bigcup\limits_{x\in S_i} N(x)$. Here, $N(x)$ is the neighborhood of node $x$ as defined in Section \ref{sec:model}. Hence, for a given schedule $(S_1,S_2,\cdots,S_k)$ where $k$ is the total number of time slots, the average detection (isolation) measure is simply $(1/k)\sum\limits_{i=1}^{k} |N(S_i)|$. 
Thus, given a bi-partite graph $\mathcal{G}(\mathcal{X}\cup\mathcal{Y},\mathcal{E})$, network life in terms of $k$ time slots, and battery supply constraint in terms of $\sigma$ time slots, the problem of finding an optimal schedule that maximizes the average detection (isolation) measure as described in Section~\ref{sec:Problem} becomes equivalent to finding a set of $k$ subsets $\{S_1,S_2,\cdots, S_k\}$, where $S_i\subseteq \mathcal{X}$, such that 

\begin{equation}
\label{eq:newProb}
\max\limits_{\{S_1,\cdots,S_k\}} {\sum\limits_{j=1}^{k} |N(S_j)|},
\end{equation}

\noindent
and each node $x\in \mathcal{X}$ is included in at most $\sigma$ such subsets.


The above problem can be cast as a graph labeling problem as described below.

\noindent
\textit{Graph Labeling Problem:}
Let $\mathcal{K}=\{1,2,\cdots,k\}$ be the set of labels, and $\mathcal{L}$ be the set of all $\sigma$-subsets\footnote{The cardinality of each subset is $\sigma$, where $\sigma$ is some positive integer.} of $\mathcal{L}$. Note that $|\mathcal{L}| = \dbinom{k}{\sigma}$. Moreover, we define

\begin{equation}
\label{eq:labeling1}
f:\;\mathcal{X}\;\longrightarrow\;\mathcal{L}
\end{equation}
i.e., $f$ is a set function that assigns $s\in\mathcal{L}$ to each vertex in $\mathcal{X}$, or in other words assign a subset of $\sigma$ labels from $\mathcal{K}$ to each $x\in\mathcal{X}$. Also, for $y\in\mathcal{Y}$, we define $F(y)$ as follows:

\begin{equation}
\label{eq:labeling2}
F(y) \triangleq \bigcup\limits_{x\in N(y)} f(x).
\end{equation}

Note that $|F(y)|$ is simply the number of distinct labels available in the neighborhood of $y$. The objective is to obtain an assignment of labels to the nodes in 
$\mathcal{X}$ (i.e., (\ref{eq:labeling1})) such that 

\begin{equation}
\label{eq:labelin3}
\text{Objective:}\;\;\; \max\limits_{f}\sum\limits_{y\in\mathcal{Y}} |F(y)|
\end{equation}
Here, the objective is to assign $\sigma$ labels to each node in $\mathcal{X}$ such that the sum of the number of distinct labels available in the neighborhood of $y$, $\forall y\in\mathcal{Y}$, is maximized. 
The scheduling problem in (\ref{eq:newProb}) and Section \ref{sec:Problem}, is equivalent to the graph labeling problem described above. 

\begin{prop}
The problem of obtaining an optimal schedule that maximizes the average detection (isolation) measures of a set of monitoring devices with limited battery supplies that cover a set of targets (target-pairs) for a given network lifetime, which is divided into $k$ time slots, is equivalent to the graph labeling problem as defined in Equations (\ref{eq:labeling1})--(\ref{eq:labelin3}).
\end{prop}

\textit{Proof --} In the graph labeling problem, let the subset of labels assigned to the vertex $x$, i.e., $f(x)\in\mathcal{L}$, correspond to the indices of time slots in which the monitoring device corresponding to $x$ is active. Since $x$ has at most $\sigma$ distinct labels by the definition of $f$, the monitoring device corresponding to node $x$ can be active in at most $\sigma$ time slots. Hence, the battery supply condition that requires a  monitoring device to be active in at most $\sigma$ time slots, is always satisfied. Moreover, $F(y)$ indicates time slots in which the target (target-pair) $y\in \mathcal{Y}$ remains covered by some $x\in\mathcal{X}$. Then, $(1/k)\sum\limits_{y\in \mathcal{Y}}|F(y)|$ is simply the average detection (isolation) measure. The set of vertices that have label $i$ correspond to the monitoring devices active in the $i^{th}$ time slot, i.e., $S_i$. Thus, finding a labeling (\ref{eq:labeling1}) that maximizes (\ref{eq:labelin3}) is basically finding a schedule $(S_1,S_2,\cdots,k)$ that maximizes the average detection (isolation) measure. \qed


An illustration of the graph labeling for the scheduling problem is given below.

\subsubsection*{Example}


In Figure \ref{fig:labeling}, instances of optimal labeling of graphs in Figures \ref{fig:example}(b) and 1(c) are shown for $\mathcal{K}=\{1,2,\cdots,5\}$ and $\sigma = 2$. Here $|\mathcal{K}|=5$ means that the given network lifetime spans five time slots. Each node $x$ has at most two labels, which represents that owing to battery constraint, a node can be active in at most two of the time slots. The node labels indicate time slots in which they remain active, thus, giving us optimal schedules. Here, the optimal detection score is 0.75, which could be obtained with the schedule $S_1=S_4=\{2\},S_2=\{4\}, S_3 = \{1,4\}, S_5=\{1\}$. Similarly, the optimal isolation score is 0.633, which could be obtained with the schedule $S_1=\{2,4\}, S_2=\{1\}, S_3=\{4\}, S_4=\{1\}, S_5=\{2\}$.

\begin{figure}[htb]\begin{center}
\includegraphics[scale=0.6]{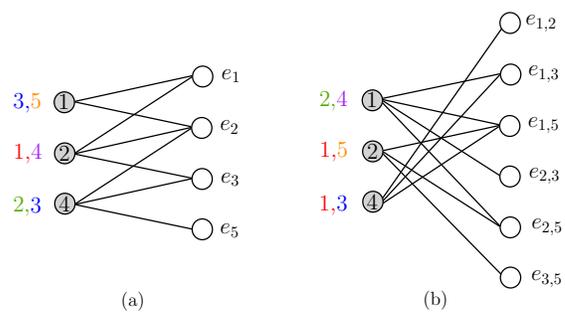}
\caption{Graph labelings for $\mathcal{K}=\{1,2,\cdots,5\}$ and $\sigma = 2$. Node labels, i.e., $f(x)$ are shown in colors.} 
\label{fig:labeling}\end{center}\end{figure}

%% file: Game_Theoretic.tex
\section{Solutions to the Graph Labeling Problem}
\label{sec:game_sol}
In this section, we provide two solution approaches to the graph labeling problem. The first one is a simple greedy heuristic, whereas, in the second approach, we utilize game-theoretic concepts. The greedy heuristic runs in polynomial time, and gives a near optimal solution for many practical networks as illustrated in the next section. However, in general, the approximation ratio of the algorithm is not known. On the other hand, the game-theoretic solution returns a graph labeling that is globally optimal with high probability.

\subsection{Greedy Heuristic}
The  graph labeling problem closely resembles the set covering problem, since we have to `cover' the set of targets using a set of monitoring nodes, each of which could cover a given subset of the targets. Since the straightforward greedy algorithm is known to be an efficient approximation algorithm for the set covering problem, we can expect it to perform well for the graph labeling problem also. Hence, we formulate a simple greedy heuristic for the graph labeling problem as follows (Algorithm \ref{algo:greedy}): For a given labeling set $\mathcal{K}$ and $\sigma$, iteratively select a combination of a label in $\mathcal{K}$ and a source node in $\mathcal{X}$ that maximizes the sum of number of distinct labels available in the neighborhoods of all target nodes in $\mathcal{Y}$. Note that in each iteration, only a source node with less than $\sigma$ labels could be selected. 

\begin{algorithm}
\caption{Greedy Heuristic}\label{algo:greedy}
\begin{algorithmic}[1]
\State \textbf{Given:} $\sigma$, $\mathcal{K} = \{1,2,\cdots,k\}$
\State \textbf{Initialization:} $\mathcal{X}'\gets \mathcal{X}$, $f(x) \gets \emptyset, \; \forall x\in\mathcal{X}$ 
\State \textbf{While} $|\mathcal{X}'|\ne\emptyset$ \textbf{do}
\State \hspace{0.25in} $(x,\ell)\gets \argmax\limits_{x\in\mathcal{X}',\ell\in\mathcal{K}}\;\; \sum\limits_{y\in\mathcal{Y}}|f(y)|$
\State \hspace{0.25in} $f(x)\gets f(x)\cup \{\ell\}$
\State \hspace{0.25in} \textbf{If} $|f(x)|=\sigma$ \textbf{do}
\State \hspace{0.45in} $\mathcal{X}'\gets \mathcal{X}'\setminus \{x\}$
\State \hspace{0.25in} \textbf{End If}
\State \textbf{End While}
\end{algorithmic}
\end{algorithm} 

If $n$ is the total number of source nodes, $m$ be the number of target nodes, and $k$ be the total number of labels in the labeling set, then greedy heuristic could be executed in at most $O(\sigma kn^2m)$ time as there are $O(\sigma n)$ iterations and each iteration could take $O(knm)$ time. Greedy heuristic gives a simple strategy to solve the labeling problem, however, we do not know the quality of the solution returned by it, that is, how far is the greedy solution from the optimal one. Therefore, we present a game-theoretic solution by posing the labeling problem as a potential game, for which algorithms are known that maintain globally optimal solution with high probability as time goes to infinity, as discussed below.

\subsection{Game Theoretic Solution to the Graph Labeling Problem}
\label{sec:Game_Th}
Game theory concepts have been extensively employed to solve locational optimization problems, such as maximizing coverage on graphs (e.g., \cite{yazicioglu2013game,zhu2013distributed}) and distributed control of multiagent systems (e.g., \cite{arslan2007autonomous,menache2011network}). In a particular approach, the idea is to determine a \textit{potential function} that captures the overall global objective. The players' individual utility functions are then appropriately aligned with the global objective such that the change in the utility of the player as a result of unilateral change in strategy equals the change in the global utility represented by the potential function. The players' strategies are then designed to ensure that local actions lead to the global objective. It turns out that this problem formulation and design can be realized using a class of non-cooperative games known as \textit{potential games}, which are now extensively used for various distributed control optimization problems.

A finite strategic game $\Gamma(\mathcal{P},\mathcal{A},\mathcal{U})$ consists of a set of players $\mathcal{P}=\{1,2,\cdots,n\}$, action space $\mathcal{A}=\mathcal{A}_1\times \mathcal{A}_2\times \cdots \times \mathcal{A}_n$ where $\mathcal{A}_x$ is a finite action set of the player $x\in \mathcal{P}$, and a set of utility functions $\mathcal{U} = \{\mathcal{U}_1,\mathcal{U}_2,\cdots,\mathcal{U}_n\}$ where $\mathcal{U}_x: \mathcal{A}\rightarrow \mathbb{R}$ is a utility function of the player $x$. If $a=(a_1,\cdots,a_x,\cdots,a_n)\in \mathcal{A}$ denotes the joint action profile, we let $a_{-x}$ denote the action of players other than the player $x$. Using this notation, we can also represent $a$ as $(a_x,a_{-x})$. 

A game is a \textit{potential game} if there exists a \textit{potential function}, $\phi: \mathcal{A}\rightarrow \mathbb{R}$ such that the change in the utility of the player $x$ as a result of a unilateral deviation from an action profile $(a_x,a_{-x})$ to $(a_x',a_{-x})$ is equal to the corresponding change in the potential function. More precisely, for every player $x$, $a_x,a_x'\in \mathcal{A}_x$, and $a_{-x}\in \mathcal{A}_{-x}$, we get 

\begin{equation}
\label{eq:potentialgame}
\mathcal{U}_x(a_x,a_{-x}) \; - \;  \mathcal{U}_x(a_x',a_{-x})\; = \; \phi(a_x,a_{-x}) \; - \; \phi(a_x',a_{-x})
\end{equation}

 
In the case of potential games, there exist algorithms, such as log-linear learning (LLL) \cite{blume1993statistical,brock2001discrete} and binary log-linear learning (BLLL) \cite{marden2012BLLL} that could be utilized to drive the players to action profiles that maximize the potential function. These algorithms embody the notion of convergence of such games to the most efficient Nash equilibrium, particularly in scenarios where utility functions are designed to ensure that the action profiles that maximize the global objective of the system coincide with the potential function maximizers \cite{blume1993statistical,marden2012BLLL}. More precisely, in potential games, LLL and BLLL algorithms guarantee that only the joint action profiles that maximize the potential function are stochastically stable \cite{marden2012BLLL}. The LLL and BLLL are in fact, \textit{nosiy best-response} algorithms that induce a Markov chain over the action space with a unique limiting distribution that depends on the noise parameter. As the nosie parameter reduces to zero, the limiting distribution has a large part of its mass over the set of potential maximizers (see e.g., \cite{marden2012BLLL,yas2} for details).

The basic idea behind these algorithms is to have \textit{noisy} best response dynamics, in which the noise parameter allows the selection of suboptimal action occasionally by the players. The probability of selecting a suboptimal action is dependent of the pay-off difference between the optimal and suboptimal cases. Thus, formulating the graph labeling problem as a potential game would allow us to use the above mentioned learning algorithms to find the most efficient solutions to the graph labeling problem. Thus, our objective now is to design a potential game corresponding to the labeling problem on graphs, and incorporate learning algorithms for the potential games to achieve the desired labeling.

%

\subsubsection{A Potential Game for the Graph Labeling}
We design a potential game $\Gamma(\mathcal{P},\mathcal{A},\mathcal{U})$ to obtain a labeling of a graph that achieves the objective in (\ref{eq:labelin3}), thus solving the scheduling problem. In our game, the set of players is the vertex set $\mathcal{X}$ in the vertex partition ($\mathcal{V}= \mathcal{X}\cup\mathcal{Y})$ of the bipartite graph $\mathcal{G}$, i.e., $\mathcal{P}=\mathcal{X}$. For each player $x\in \mathcal{X}$, the action set $\mathcal{A}_x$ is the set of all $\sigma$-subsets of the labeling set $\mathcal{K}=\{1,\cdots,k\}$. 
We also need to have a potential function that captures the global objective. For this, we define $S_j$ as the set of vertices with the label $j$, i.e.,
\begin{equation}
\label{eq:Ix}
S_j \; = \; \{x\in \mathcal{X}: j\in f(x)\}
\end{equation}

A potential function is then defined as
\begin{equation}
\label{eq:potential}
\phi(a) \; \triangleq \; \mathlarger{‎‎\sum}_{j=1}^{k}\left\lvert \bigcup\limits_{x\in S_j} N(x)\right\rvert
\end{equation}

Note that $\phi(a)$ is simply the total number of nodes in $\mathcal{Y}$ having a label $j\in \mathcal{K}$ in their neighborhoods, summed over all the labels, which is equivalent to the $\sum\limits_{y\in \mathcal{Y}}\lvert F(y)\rvert$ in (\ref{eq:labelin3}). Thus, $\phi(a)$ indeed captures the global objective.

Moreover, we define the utility function of the player $x$ as the total number of labels made available by $a_x$ to the nodes in $N(x)$ that otherwise would not have been available to the nodes in $N(x)$. For instance, in Figure \ref{fig:labeling}(a), node $1$ has labels $\{3,5\}$, which represents the action $a_1$. Moreover, for the two neighbors of node $1$, i.e., $e_1$ and $e_2$, node $1$ is the only one with the label $5$; and for the node $e_1$, node $1$ is the only one with the label $3$. Thus, $U_1(a_1,a_{-1}) = 2+1=3$. More precisely, we define $U_x(a_x,a_{-x})$ as

\begin{equation}
\label{eq:utility}
U_x(a_x,a_{-x}) \; \triangleq \; \mathlarger{‎‎\sum}_{j=1}^{k} a_{xj}\left\lvert N(x) \setminus \bigcup_{z\in S_j\setminus \{x\}}N(z) \right\rvert 
\end{equation}
where, 
\begin{equation*}
a_{xj} = 
\left\{
\begin{array}{llll}
1 & \;\;\text{if} \;\; j\in a_{x} (=f(x))\\
0 & \;\;\text{otherwise.}
\end{array}
\right.
\end{equation*}

Next, we show that with the potential function as defined in (\ref{eq:potential}), and the utility function as in (\ref{eq:utility}), the game designed above is indeed a potential game. 

\begin{theorem}
$\Gamma(\mathcal{P},\mathcal{A},\mathcal{U})$ is a potential game if utilities are defined as in (\ref{eq:utility}).
\end{theorem}

\textit{Proof} -- The potential function, as defined in (\ref{eq:potential}) can be written as, 
\begin{equation}
\label{eq:pot1}
\begin{split}
& \phi(a_x,a_{-x})  = \mathlarger{‎‎\sum}_{j=1}^{k}\left\lvert \bigcup\limits_{x\in S_j} N(x)\right\rvert\\
& = \mathlarger{‎‎\sum}_{j=1}^{k}\left(a_{xj}\left\lvert N(x) \setminus \bigcup_{z\in S_j\setminus \{x\}}N(z) \right\rvert \; + \; \left\lvert \bigcup_{z\in S_j\setminus \{x\}}N(z) \right\rvert \right)\\
& = \mathlarger{‎‎\sum}_{j=1}^{k} a_{xj}\left\lvert N(x) \setminus \bigcup_{z\in S_j\setminus \{x\}}N(z) \right\rvert \; + \;\mathlarger{‎‎\sum}_{j=1}^{k}\left\lvert \bigcup_{k\in I_x\setminus \{i\}}N(z) \right\rvert\\
& = \mathcal{U}(a_x,a_{-x}) \; + \; \mathlarger{‎‎\sum}_{j=1}^{k}\left\lvert \bigcup_{z\in S_j\setminus \{x\}}N(z) \right\rvert
\end{split}
\end{equation}

Similarly, for $a=(a_x',a_x)$, we get
\begin{equation}
\label{eq:pot2}
\begin{split}
\phi(a_x',a_{-x})  & = \mathcal{U}(a_x',a_{-x}) \; + \; \mathlarger{‎‎\sum}_{j=1}^{k}\left\lvert \bigcup_{z\in S_j\setminus \{x\}}N(z) \right\rvert
\end{split}
\end{equation}
Subtracting (\ref{eq:pot2}) from (\ref{eq:pot1}) gives us the desired result, i.e., 
\begin{equation*}
\phi(a_i,a_{-i}) - \phi(a_i',a_{-i}) = U(a_i,a_{-i}) - U(a_i,a_{-i}) 
\end{equation*}\qed

Since our graph labeling problem can be formulated as a potential game, using the results in \cite{marden2012BLLL} we deduce that \textit{if players adhere to the binary log linear algorithm (stated below), then the objective in (\ref{eq:labelin3}) is maximized}. In other words, if $\sigma$ unique labels from a total of $k$ labels are assigned to each node $x\in\mathcal{X}$ as per below algorithm, then the number of distinct labels in the neighborhood of every node $y\in\mathcal{Y}$ is likely to converge to the maximum value.

\begin{algorithm}
\caption{Binary Log-Linear Learning \cite{marden2012BLLL}}\label{algo:BLL}
\begin{algorithmic}[1]
\State \textbf{Initialization:} Pick a small $\epsilon \in\mathbb{R}_+$, an $a\in \mathcal{A}$, and total number of iterations.
\State \textbf{While} $i\le \text{number of iterations}$ \textbf{do}
\State \hspace{0.25in} Pick a random node $x\in \mathcal{X}$, and a random $a_x'\in \mathcal{A}_x$.
\State \hspace{0.25in} Compute $P_\epsilon\; = \; \frac{\epsilon^{U_x(a_x',a_{-x}(t))}}{\epsilon^{U_x(a_x',a_{-x}(t))}\; + \; \epsilon^{U_x{(a_x,a_{-x}(t))}}}$.
\State \hspace{0.25in} Set $a_x\leftarrow a_x'$ with probability $P_\epsilon$.
\State \hspace{0.25in} $i\gets i+1$
\State \textbf{End While}
\end{algorithmic}
\end{algorithm} 

Note that initially the nodes are assigned $\sigma$-element subsets of labels randomly. Afterwards, in each iteration, a node is selected at random, and a $\sigma$-subset of labels that improve the overall labeling to attain the objective in (\ref{eq:labelin3}), is selected with a certain probability. 

%% file: New_Simultaneous_Placement.tex
\section{Simultaneous Placement and Scheduling of Monitoring Devices}
\label{sec:simultaneous}
So far, we have considered the optimal scheduling of resource bounded monitoring devices, assuming that their placement is fixed, i.e., locations at which monitoring devices are deployed are given. If $\mathcal{S}$ is the set of all such nodes at which monitoring devices could be deployed, then the \textit{placement problem} is to select a subset $\mathcal{X}\subseteq \mathcal{S}$ with the given cardinality such that the number of targets (pair-wise targets) that are covered, i.e., lie within the range of at least one such device, is maximized. Typically, to maximize the coverage of targets for a given network lifetime, the placement problem is first solved, followed by the determination of efficient schedules for the monitoring devices.

\begin{figure}[!htb]\begin{center}
\includegraphics[scale=0.6]{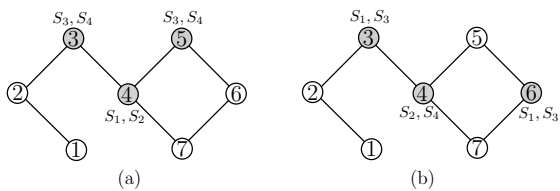}
\caption{(a) Optimal schedule for a given placement. (b) Optimal placement and schedule of three monitoring devices with $\lambda=1$, $\sigma=2$ for $k=4$.} 
\label{fig:P_S_Example}\end{center}\end{figure}

However, for a given network lifetime, and a fixed number of resource bounded monitoring devices, simultaneously optimizing their placement and scheduling maximizes the average detection (isolation) measure. For instance, consider the network in Fig. \ref{fig:P_S_Example}, in which three monitoring devices with $\lambda = 1$ and $\sigma=2$ are deployed to cover the maximum number of nodes for $k=4$. Fixing the placement of devices at nodes $\{3,4,5\}$, optimal schedule (for instance, $S_1=S_2=\{4\},S_3=S_4=\{3,5\}$) gives $\mathcal{D} = 0.642$, whereas the maximum possible $\mathcal{D}$ under the conditions is $0.714$, which could be obtained by placing the devices at nodes $\{3,4,6\}$ and with a schedule $S_1=S_3 = \{3,6\}, S_2=S_4=\{4\} $.  

The BLLL based algorithm to schedule a set of monitoring devices with fixed locations, presented in Section \ref{sec:Game_Th}, can be modified to simultaneously optimize the placement as well as scheduling of such devices to maximize the average detection (isolation) measure. This modification is presented as Algorithm \ref{algo:Mod_BLL} below. Fixing the number of monitoring devices $|\mathcal{X}|$, the objective is to \textit{select $\mathcal{X}\subseteq \mathcal{S}$, and assign at most $\sigma$ labels to each node from a labeling set $\mathcal{K} = \{1,2,\cdots,k\}$ so that the average detection measure $\mathcal{D}$ (or the isolation measure $\mathcal{I}$) is maximized}. The labeling of nodes selected in $\mathcal{X}$ will then give the schedule.


In this case, players $\mathcal{P}$ are the monitoring devices, for which we need to find the locations, i.e., the nodes at which they are deployed, as well as schedules, i.e., time slots in which they become active. Using the same notations as in Section \ref{sec:Game_Th}, here, action of a player is the selection of $(x,a_x)\in(\mathcal{S}\times \mathcal{A}_x)$, where $\mathcal{S}$ is the set of all such nodes at which a monitoring device could be placed, and $\mathcal{A}_x$ is the set of all possible $\sigma$-subsets of the labeling set $\mathcal{K}$. Previously, the choice of $x$ was fixed for a monitoring device and the player's action comprised of only selecting $a_x$. Similarly, utility of a player for the choice of an action $(x,a_x)$ here is simply the number new labels that become available in the neighborhood of node $x$ as a result of assigning labels in $a_x$ to $x$. In the search of a better solution, in each iteration, a new action $(s,a_s)$ is selected with a certain probability for a randomly selected player. It simply means that with a certain probability, either new labels are assigned to the node at which (randomly selected) player is located, or a new node as well as a new set of labels (selected at random) are chosen for the player.

\begin{algorithm}
\caption{Simultaneous Placement and Scheduling}\label{algo:Mod_BLL}
\begin{algorithmic}[1]
\State \textbf{Initialization:} Pick a small $\epsilon \in\mathbb{R}_+$ and the number of iterations. Select randomly a subset of nodes $\mathcal{X}\subseteq \mathcal{S}$, and assign labels to nodes in $\mathcal{X}$, i.e, select $a\in \mathcal{A}$.
\State \textbf{While} $i\le\text{number of iterations}$ \textbf{do}
\State \hspace{0.25in} Randomly select a node $x\in \mathcal{X}$.
\State \hspace{0.25in} Randomly select a node $s\in (\mathcal{S}\setminus\mathcal{X})\cup\{x\}$, and $a_s\in\mathcal{A}_s$.
\State \hspace{0.25in} Compute $P_\epsilon\; = \; \frac{\epsilon^{U_s(a_s,a_{-x})}}{\epsilon^{U_s(a_s,a_{-x})}\; + \; \epsilon^{U_x{(a_x,a_{-x})}}}$.
\State \hspace{0.25in} With probability $P_\epsilon$, set $\mathcal{X}\gets \left(\mathcal{X}\setminus\{x\} \right)\cup\{s\}$, and select $a_s$ for node $s$.
\State \hspace{0.25in} $i\gets i+1$
\State \textbf{End While}
\end{algorithmic}
\end{algorithm} 

Simulation results for the above algorithm are illustrated in Section \ref{sec:sim_simultaneous}. Using various networks, it is shown that simultaneously selecting the locations for monitoring devices as well as scheduling them using Algorithm \ref{algo:Mod_BLL}, gives improved average detection as compared to the one obtained by solving the placement and scheduling separately. 

%% file: Simulations.tex
\section{Numerical Results}
\label{sec:simulations}
\begin{figure*}[!htb]\begin{center}
\includegraphics[scale=0.45]{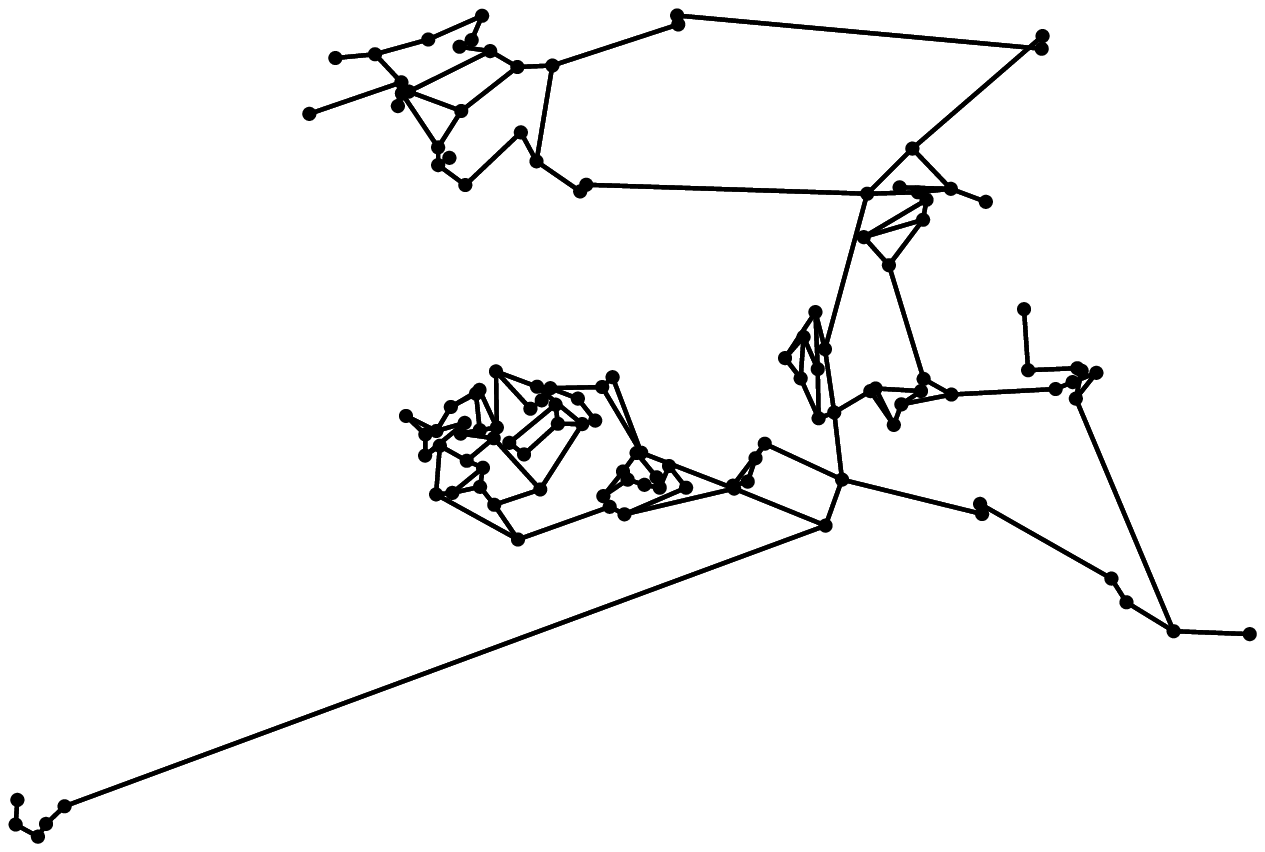}
\includegraphics[width=2.9in,height=2.15in]{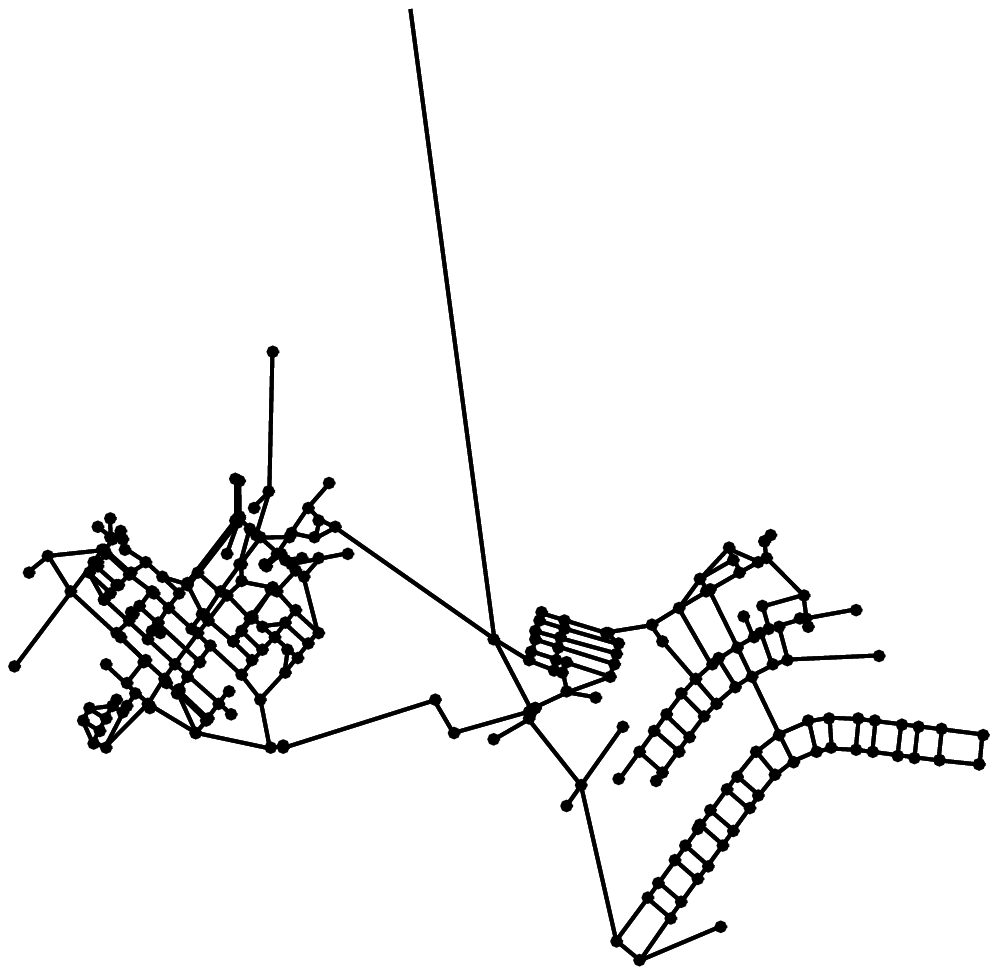}
\caption{Schematics for water networks 1 and 2.} 
\label{fig:layout_1}\end{center}\end{figure*}


\begin{figure*}[!htb]\begin{center}
\includegraphics[scale=0.4]{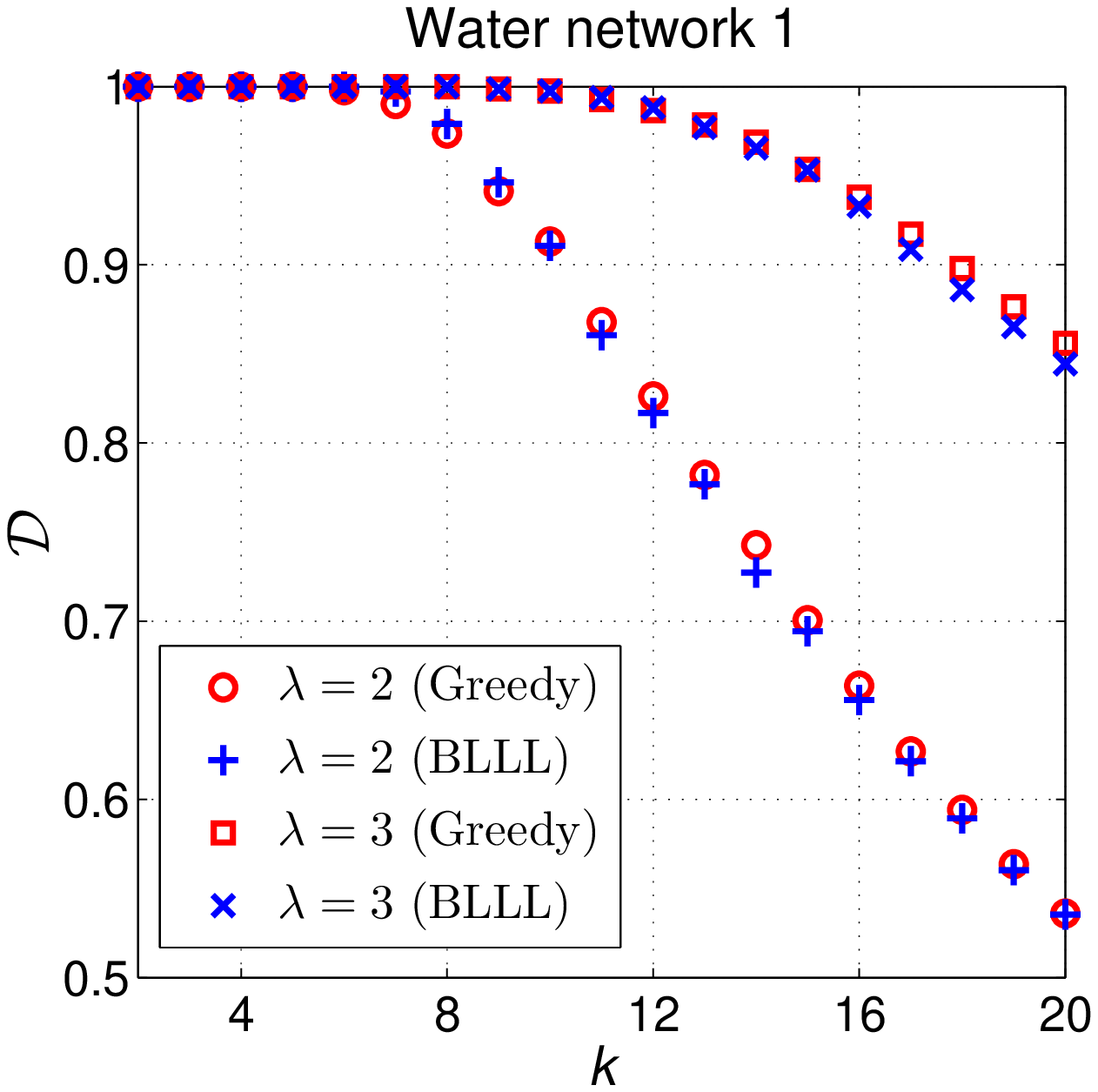}
\includegraphics[scale=0.4]{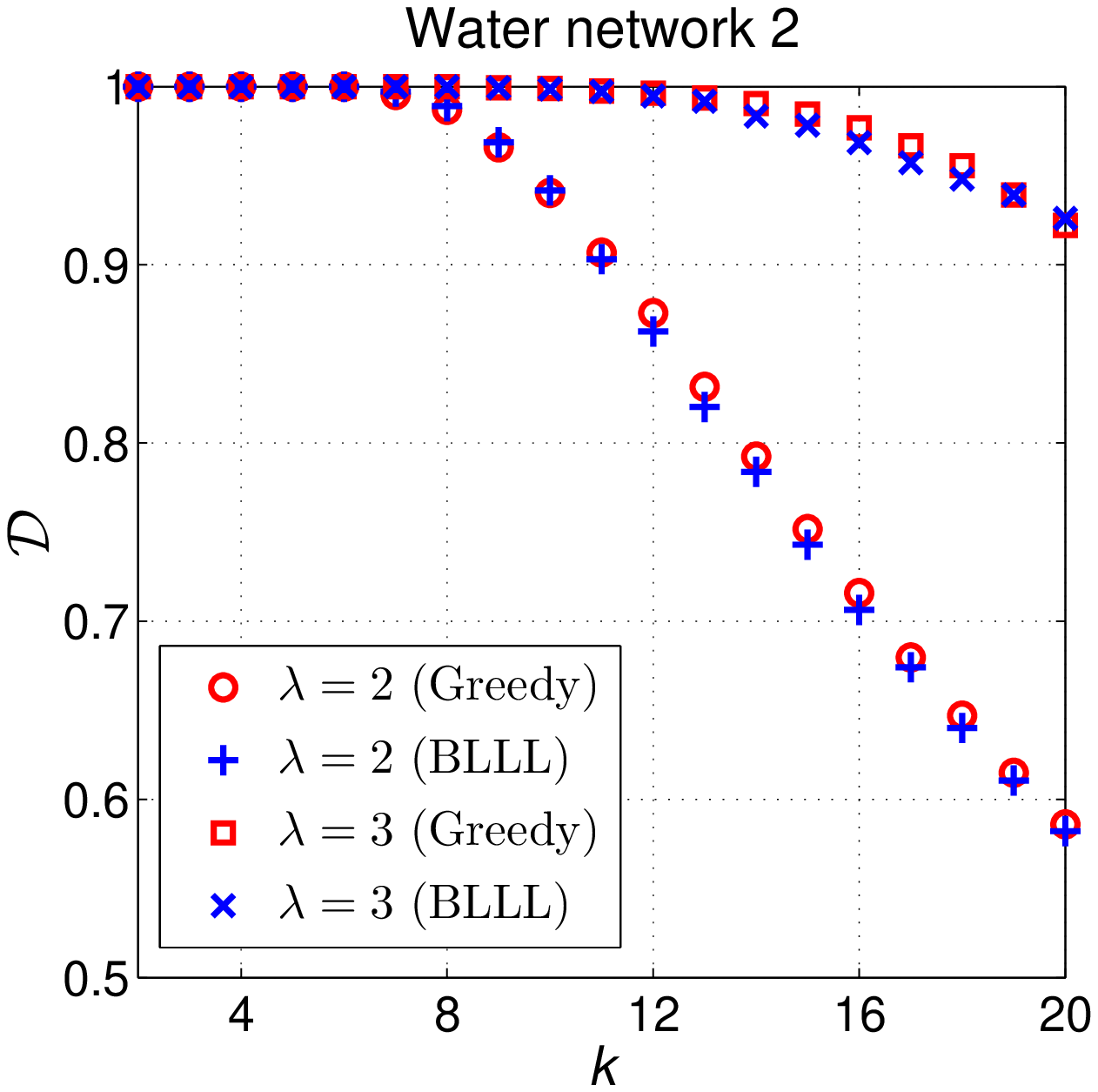}
\includegraphics[scale=0.4]{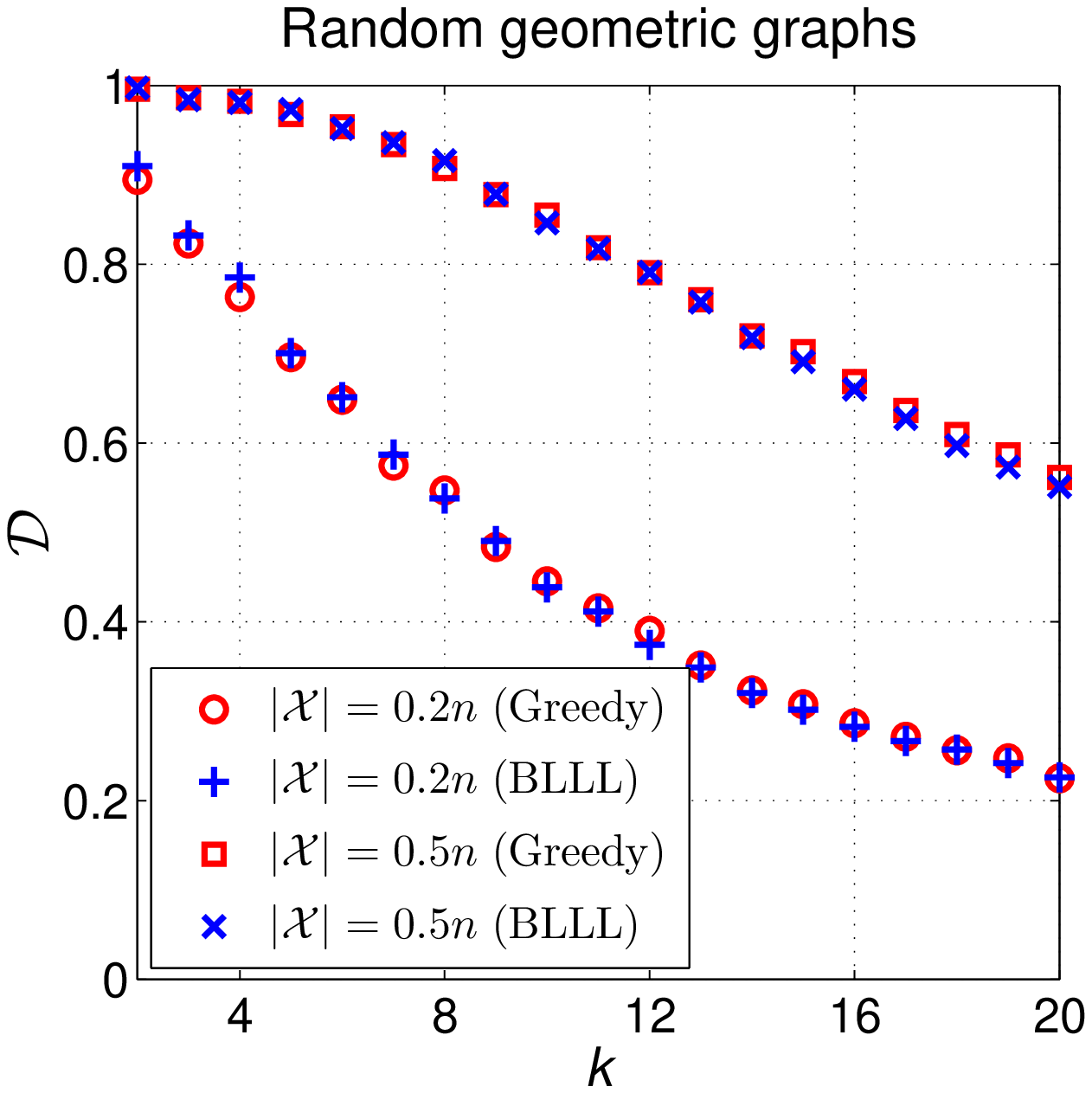}
\caption{Plots of $\mathcal{D}$ as a function of network lifetime $k$ for scheduling on water networks and random geometric networks, assuming that each monitoring device has a battery lifetime of $\sigma=2$ time slots.} 
\label{fig:plots}\end{center}\end{figure*}

\begin{figure*}[!htb]\begin{center}
\includegraphics[scale=0.32]{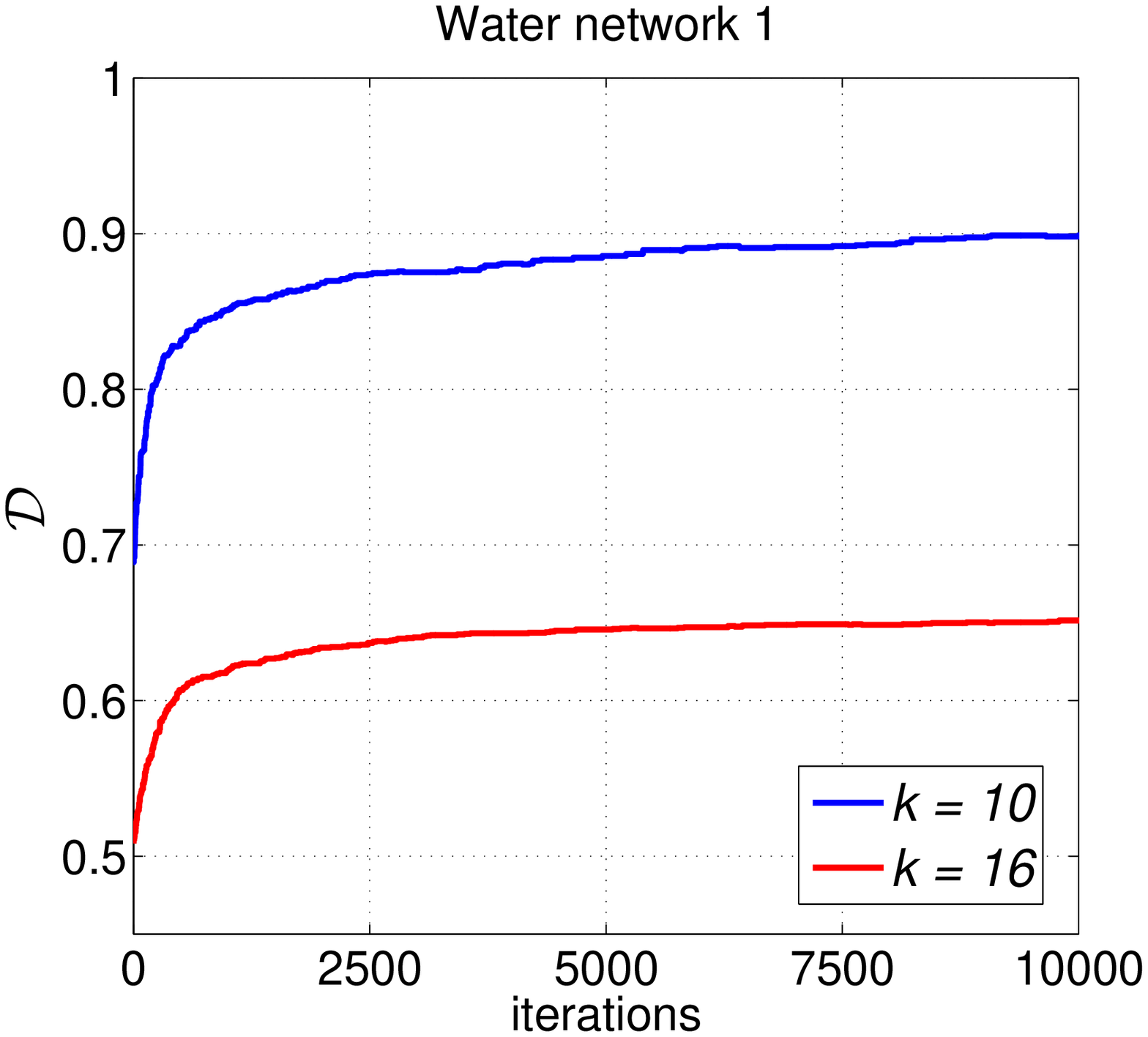}
\includegraphics[scale=0.32]{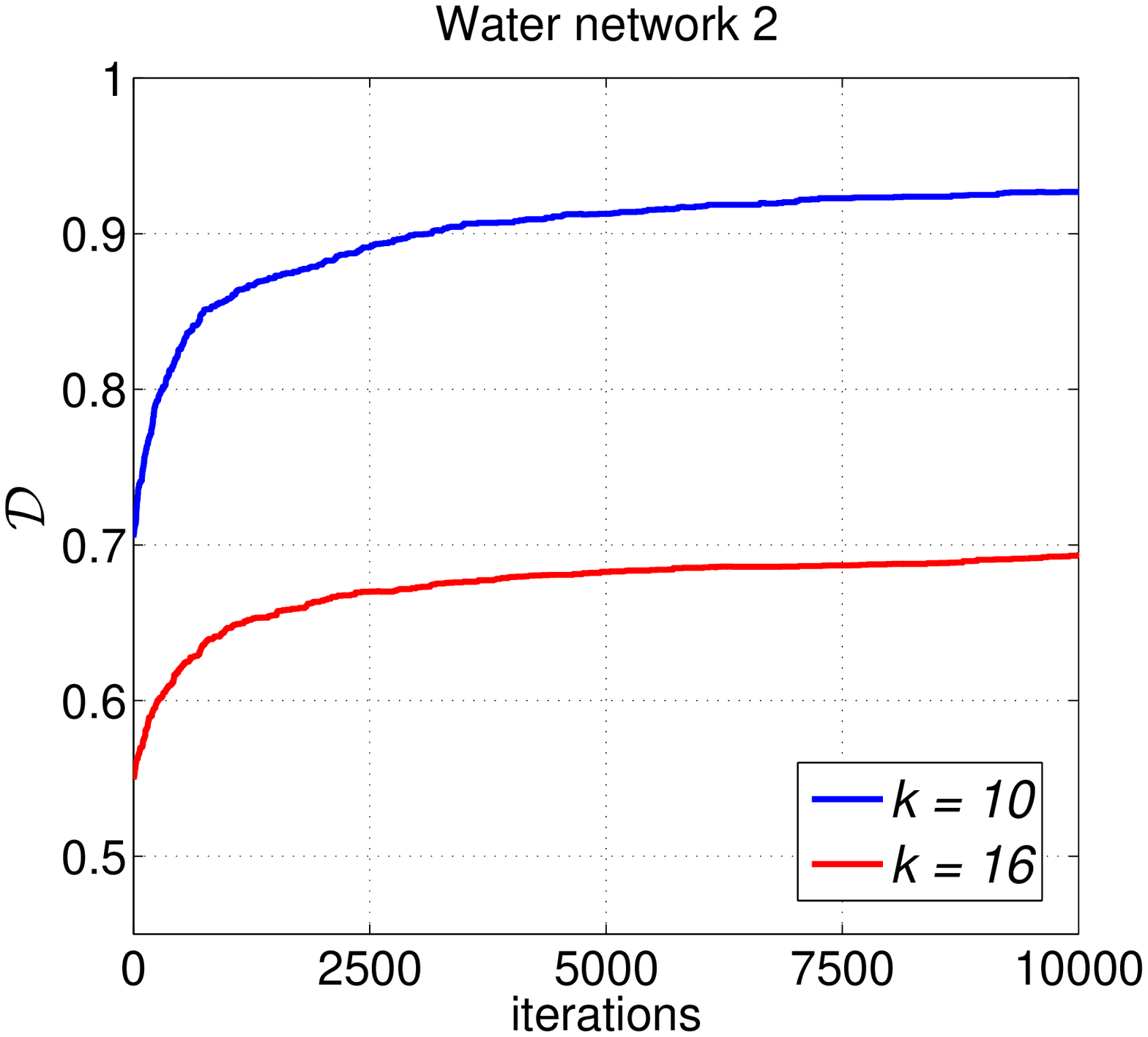}
\includegraphics[scale=0.32]{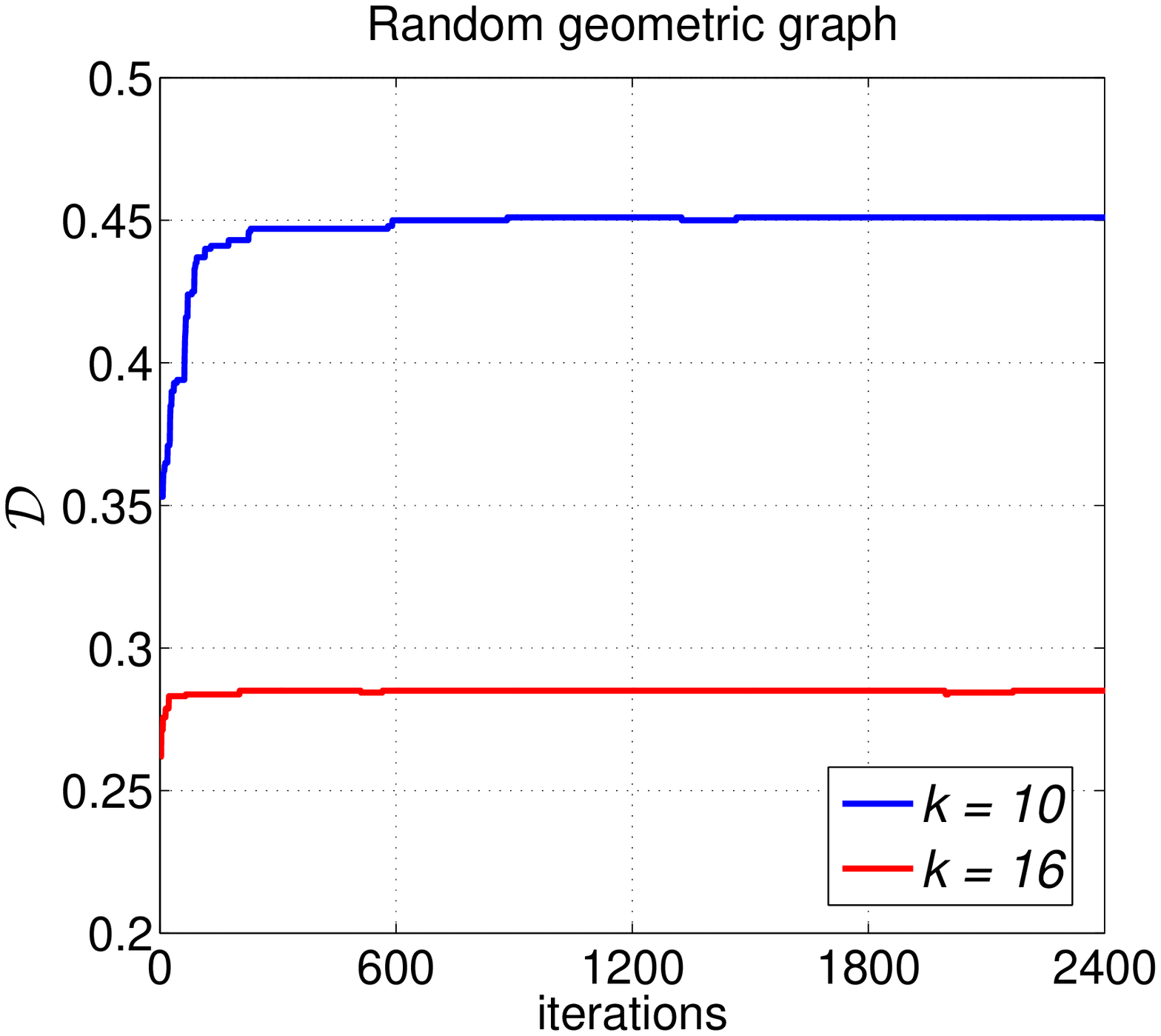}
\caption{Plots of $\mathcal{D}$ as a function of (BLLL) iterations to illustrate the convergence of BLLL algorithm for the scheduling of monitoring devices with $\sigma=2$ and $k=10,16$.} 
\label{fig:convergence}\end{center}\end{figure*}

In this section, we present numerical results on the simple greedy and BLLL based algorithms for the scheduling and placement of monitoring devices on urban water distribution networks and random geometric networks as explained below.
\subsection{Scheduling Monitoring Devices in in Water Distribution Networks}
Water distribution networks can be modeled as undirected graphs in which edges represent the pipes and nodes represent the junctions. To detect pipe bursts and leakages, pressure sensors are deployed at junctions, which could sense the pressure transient generated as a result of pipe burst within a certain distance (range) from the sensor. The distance threshold based model has been used in water networks in the context of sensor placement problems, e.g., \cite{deshpande2013optimal,abbas2015efficient}. The pressure sensors are battery operated devices with limited battery lifetime. Thus, top operate these sensors for an extended period of time, they need to be scheduled. Here, we simulate scheduling algorithms, including simple greedy and BLLL based algorithm for the efficient scheduling of monitoring devices, which are pressure sensors in this case, to obtain high values of $\mathcal{D}$ in two different water distribution networks. The details of these networks, referred to as the \textit{Water Network 1} and \textit{Water Network 2}, are as follows:

Water Network 1 \cite{ostfeld2008battle,WN1} has 126 nodes, 168 pipes, one reservoir, one pump, and two storage tanks. This benchmark water distribution network has been extensively studied in the context of sensor placement problems for water quality. Water network 2 \cite{jolly2013research} is a grid system in Kentucky with 366 pipes, 270 nodes, three tanks, and five pumps. The layouts of both networks are illustrated in Figure \ref{fig:layout_1}. For both the networks, we consider that the sensors are deployed at the junctions as source nodes (monitoring devices), i.e., $\mathcal{X}$, and the set of pipes, which are edges in the corresponding network graph, as targets, i.e., $\mathcal{Y}$. Moreover, for each sensing device, we assume $\sigma = 2$, and compute $\mathcal{D}$ for a network lifetime, given by $k$ time slots, using greedy and BLLL algorithms. For each BLLL instance, we perform 20,000 iterations by selecting $\epsilon$ to be $0.015$. The plots of $\mathcal{D}$ as a function of $k$ for various ranges of sensing devices (as defined in Section \ref{sec:model}) are given in Figure \ref{fig:plots}. 

We can see that both greedy and BLLL gives approximately same results. However, BLLL has an advantage over the greedy algorithm as it allows to simulatneously solve the placement as well as scheduling problem (as discussed in Section \ref{sec:simultaneous}), which gives improved $\mathcal{D}$ as compared to individually solving placement and scheduling problems. Moreover, if BLLL is run for sufficiently large number of iterations, the algorithm converges to the optimal solution. Similar plots can be obtained for the scheduling of monitoring devices to maximize the average isolation measure $\mathcal{I}$ by first obtaining the appropriate network representation as outlined in Section \ref{sec:labeling_A}. In Figure \ref{fig:convergence}, the convergence of BLLL algorithm is illustrated. For both water networks, $\mathcal{D}$ as a function of iterations is shown for $\lambda=2$, $\sigma=2$, and $k=10$ and $16$. We observe the algorithm converges to near optimal value fast, within about 5000 iterations, and the improvements thereafter, are quite small.

\subsection{Scheduling Monitoring Devices in Random Geometric Networks}
Random geometric networks are a form of spatial networks in which nodes are deployed uniformly at random in a certain area. An edge exists between two nodes if the Euclidean distance between them is at most $r$, which is often referred to as radius of the sensing footprint. Owing to a wide variety of applications in various domains, such as modeling of wireless sensor networks, these networks have been extensively studied. For our simulations, we consider a network with $100$ nodes, deployed uniformly at random over an area of $10\times 10$, and $r=2$. The set of targets here is the set of all nodes. Moreover, a certain fraction of nodes (either $20\%$ or $50\%$) are selected at random as source nodes, i.e., nodes with monitoring devices. A monitoring device has a battery lifetime of at most $\sigma=2$ time slots, and can monitor targets that are at a Euclidean distance of at most $2$ from it.\footnote{In terms of the (graph) distances as defined in Section \ref{sec:model}, the range of each monitoring device is $\lambda = 1$, as the Euclidean distance of at most 2 between two nodes $u$ and $v$ implies $d(u,v)=1$.} In Figure \ref{fig:plots}, $\mathcal{D}$ as functions of $k$ are illustrated using greedy and BLLL algorithms. Each point on the plots is an average of fifty randomly generated graph instances. In Figure \ref{fig:convergence}, the convergence of BLLL algorithm is shown for some instances of random geometric graphs with 100 nodes, out of which 20 randomly selected nodes contain monitoring devices.
\subsubsection{Random Scheduling in Random Networks}
Another special case of interest is related to the quality of random scheduling, i.e., given a total of $k$ time slots, if each node remains active in $\sigma$ time slots chosen randomly, then what is the average detection performance of such a random scheduling? Here, we discuss this question for random networks, including random geometric networks and networks that could be modeled by Erd\H{o}s-R\'{e}nyi random graphs. Though random scheduling is inferior to the BLLL based scheduling in terms of the detection (or isolation) performance, it is useful in many scenarios since it neither requires any sort information regarding the network structure, nor requires any coordination between the monitoring devices. The average detection measure of random scheduling in random geometric networks is given below.

\begin{prop}
\label{thm:random}
Let $G(V,E)$ be a random geometric graph in which each node contains a monitoring device that remains active in $\sigma$ time slots that are randomly chosen from a total of $k$ time slots, which correspond to the overall lifetime of the network. If each node in a graph is also a target, then the average detection performance of this random scheduling is 
\begin{equation}
\label{eq:rand_sch_ER}
\mathcal{D}(G) = 1 - \frac{(k-\sigma)}{k}\text{exp}\left(\frac{-\sigma\lambda\pi r^2}{k}\right)
\end{equation}
where $r$ is the radius of the sensing footprint of node, and $\lambda$ is the number of nodes per unit area.
\end{prop}


A proof of the above theorem is given in the Appendix. As above, it can be shown that in the case of Erd\H{o}s-R\'{e}nyi random graphs with $n$ nodes, denoted by $G_{n,p}$, in which any two nodes are adjacent with some probability $p$, this random scheduling scheme results in an average detection performance given by

\begin{equation}
\label{eq:ER_Random}
\mathcal{D}(G_{n,p}) = 1 - \frac{(k-\sigma)}{k}\text{exp}\left(\frac{-\sigma}{k}np\right) 
\end{equation}

Note that in (\ref{eq:ER_Random}), it is assumed that all the nodes have monitoring devices and all the nodes need to be covered.


\subsection{Simultaneous Placement and Scheduling of Monitoring Devices Using Algorithm \ref{algo:Mod_BLL}}
\label{sec:sim_simultaneous}
We illustrate the Algorithm \ref{algo:Mod_BLL} for the water network 1 and the random geometric graph here. For the water network 1, we set the number of monitoring devices to be $25$, where each device has a range $\lambda = 2$. The set of pipes (or edges in the corresponding network graph) are the targets that need to be covered by these devices. We simulate two scenarios; in the first case we use Algorithm \ref{algo:Mod_BLL} to simultaneously select the nodes and schedules for the monitoring devices; in the second scenario, we first solve the placement problem by selecting the 25 nodes, say $\mathcal{X}\subset V$, that maximize the number of edges that are at most distance 2 from some node in $\mathcal{X}$, and then solving the scheduling problem using Algorithm \ref{algo:BLL}. We note here that the placement problems, in this context, are typically solved using some variant of the minimum set cover problem, or the maximum coverage problem in case the number of monitoring devices is fixed (e.g., \cite{perelman2015sensor,krause2008efficient,krysander2008sensor}). Since the number of devices is fixed here, and the targets to be covered are edges, we use the maximum coverage problem to place (a given number of) monitoring devices at nodes that maximize the number of edges that are at most $\lambda=2$ distance from at least one of the selected nodes. Moreover, since maximum coverage problem is NP-hard, we use a greedy heuristic, which gives best approximation ratio, to solve it.

\begin{figure}[!htb]\begin{center}
\includegraphics[scale=0.32]{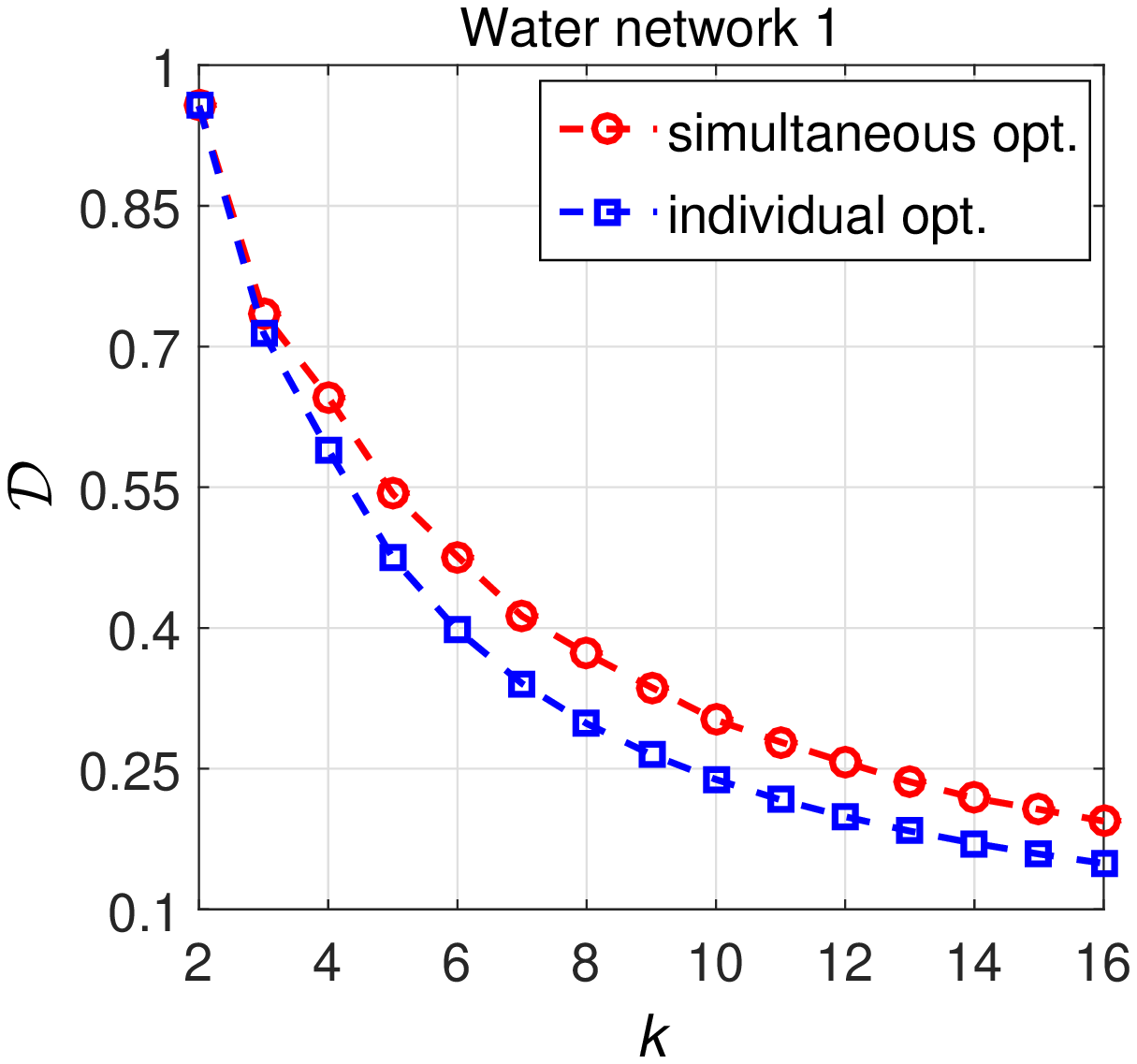}
\includegraphics[scale=0.32]{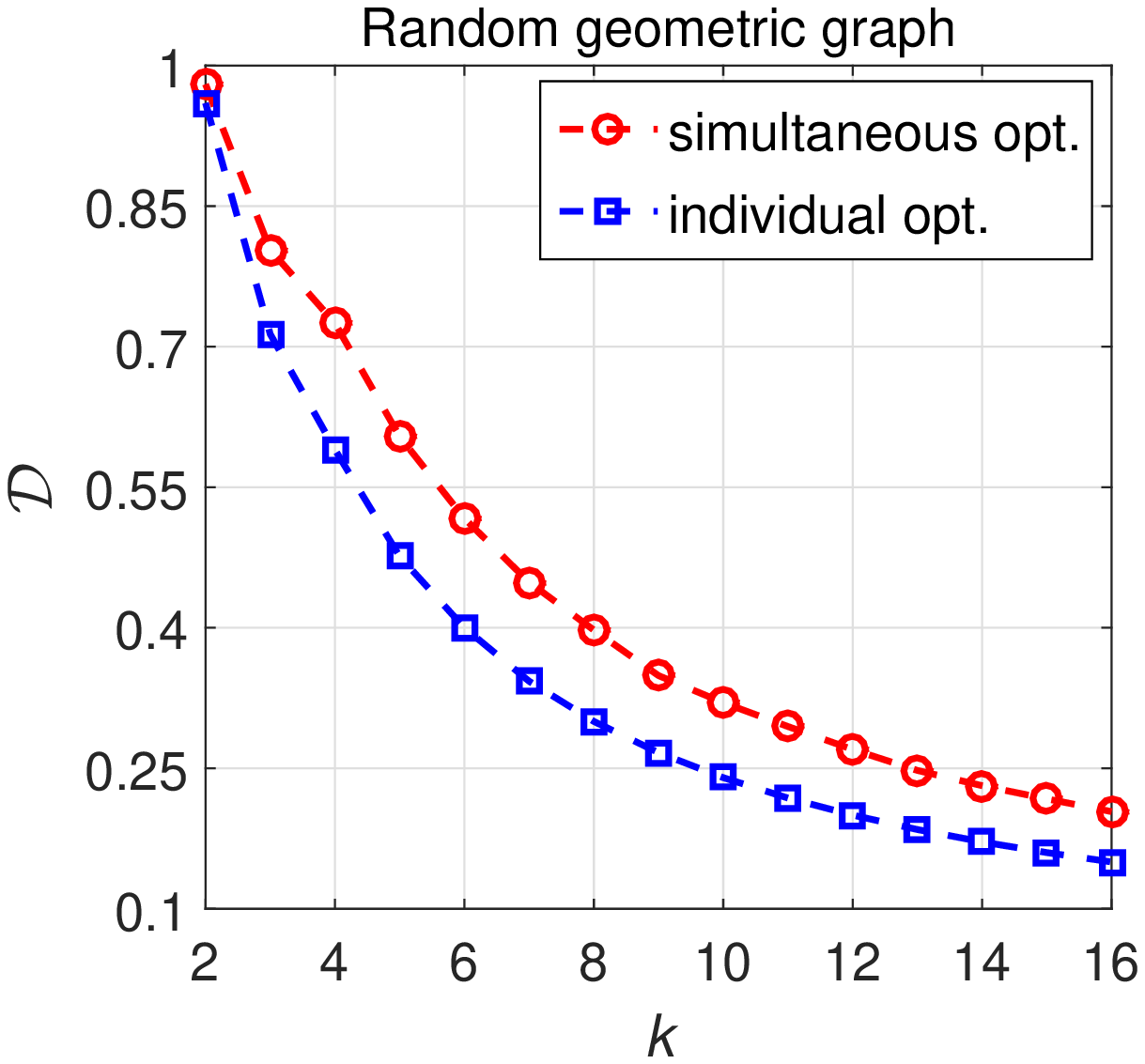}
\caption{Comparison of simultaneously optimizing scheduling and placement using Algorithm \ref{algo:Mod_BLL} versus separately optimizing placement and scheduling in terms of $\mathcal{D}$ as a function of $k$.} 
\label{fig:sim}\end{center}\end{figure}

\begin{figure}[!htb]\begin{center}
\includegraphics[scale=0.3]{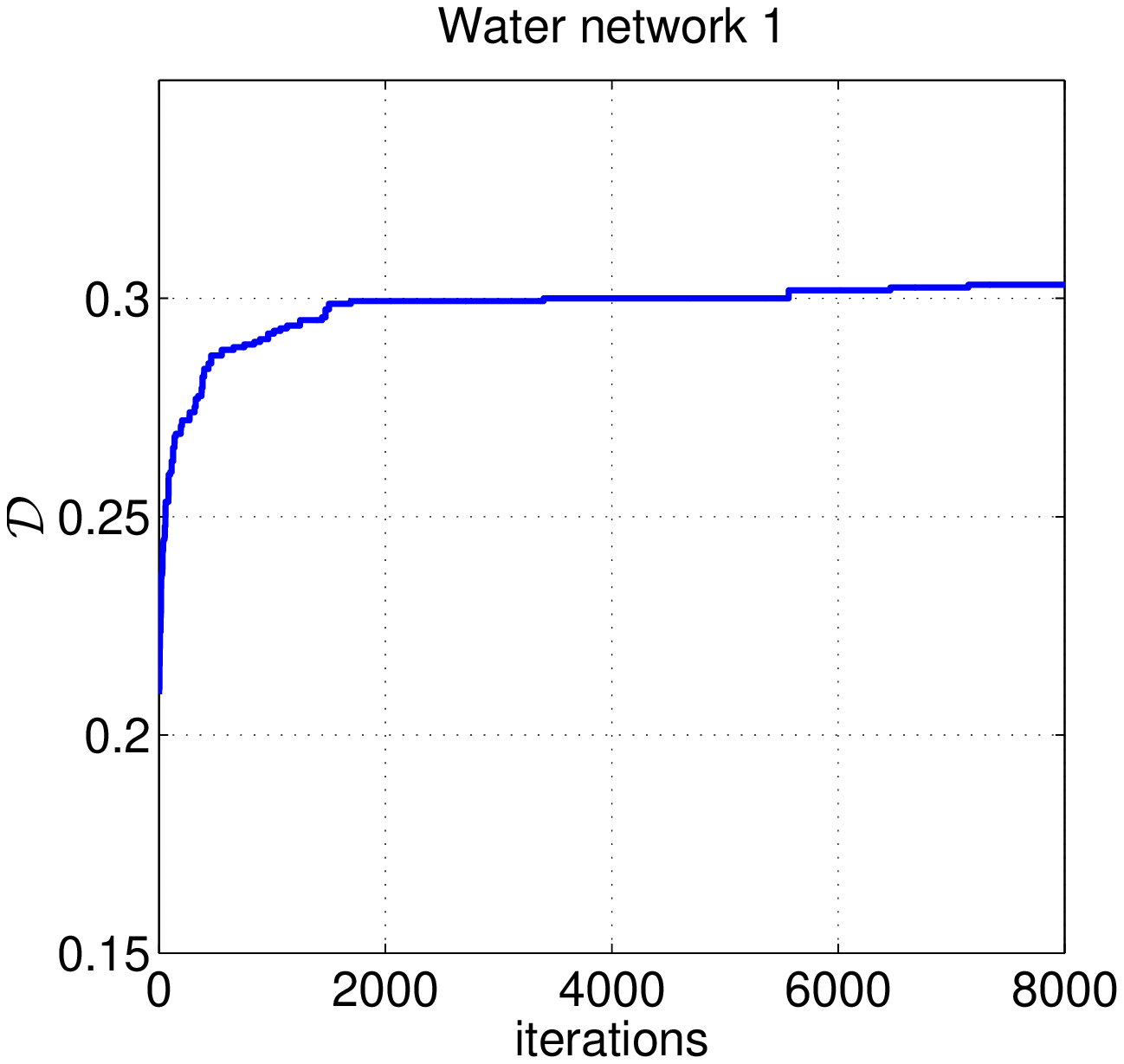}
\includegraphics[scale=0.315]{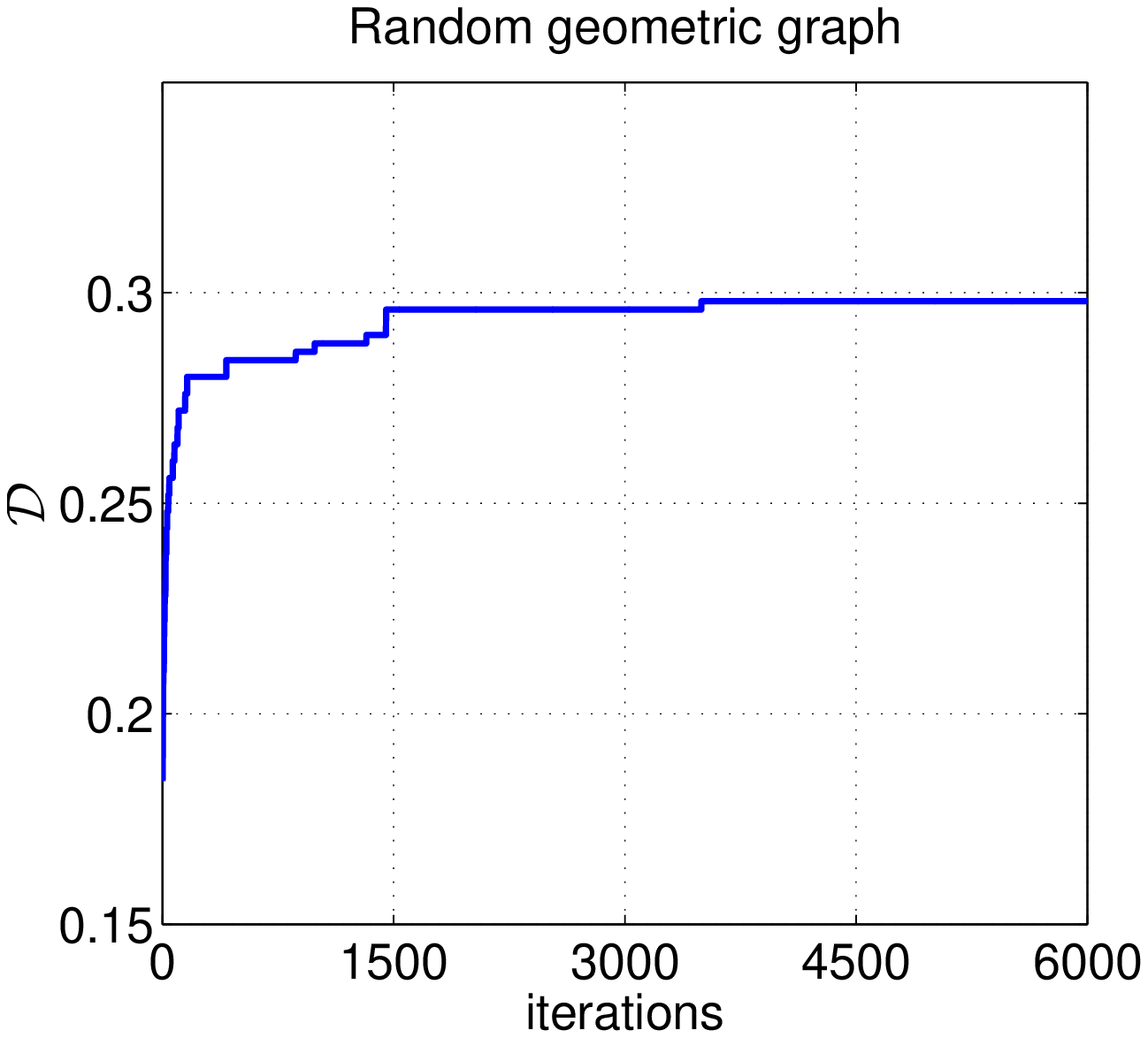}
\caption{Plots of $\mathcal{D}$ as a function of iterations of Algorithm \ref{algo:Mod_BLL} showing the convergence of the algorithm.} 
\label{fig:sim_iter}\end{center}\end{figure}

The results are illustrated in Figure \ref{fig:sim}. It can be seen that Algorithm \ref{algo:Mod_BLL} (simultaneously solving placement and scheduling) is always giving higher average detection $\mathcal{D}$. For the random geometric graph, we simulate instances consisting of 50 nodes deployed at random in an area of $500\times 500$ $\text{unit}^2$, out of which 10 could contain monitoring devices capable of covering nodes within a Euclidean distance of 100 units. The targets here are nodes, and the objective is to maximize the average detection for a given network lifetime. As with the water network example, average detection is improved if placement and scheduling is solved simultaneously using Algorithm \ref{algo:Mod_BLL} as compared to optimizing placement and scheduling separately. For all cases, the battery lifetime of each monitoring device is assumed to be $\sigma = 2$ time slots. In Figure \ref{fig:sim_iter}, we illustrate the convergence of Algorithm \ref{algo:Mod_BLL} for the water network 1 and random geometric graph example. given the network lifetime $k=10$ time slots.
 
%% file: Special_Cases.tex
\section{Special Case: Scheduling to Maximize Network Lifetime While Ensuring Complete Coverage}
\label{sec:special}
%
An important special case of the scheduling problem is to control the activity of monitoring devices such that the overall network lifetime is maximized while ensuring \textit{complete coverage}, i.e., $\mathcal{D} = 1$. In a basic setting, we consider that all nodes in a graph need to be covered at all times, and each node is equipped with a monitoring device that can remain active in at most $\sigma$ time slots. Then, the objective is to schedule these monitoring devices such that the number of time slots $k$, in which all of the nodes remain covered through a subset of active devices, is maximized.

The problem is related to the notion of \textit{dominating sets in graphs}. 

\begin{definition}
\label{def:dominating}
A dominating set is a subset of vertices in a graph $S_i\subseteq V$, such that for every $u\in V$, either $u\in S_i$, or there exists some $v\in S_i$ such that $v\in N(u)$.
\end{definition}

In other words, considering the targets to be the set of nodes (i.e., $\mathcal{Y}=V$), the ranges of monitoring devices $\lambda$ to be 1, the network is guaranteed to be completely covered whenever the set of nodes with active monitoring devices form a dominating set in the network graph. Moreover, in the case of targets being edges (i.e., $\mathcal{Y}=E$), a dominating set of active monitoring devices with ranges $\lambda=2$ is also sufficient for the complete coverage of targets within the network. Thus, to maximize the overall network lifetime while ensuring complete coverage of targets, the problem of finding distinct dominating sets in a graph is of great importance. The problem of finding distinct dominating sets under certain constraints has been of great interest owing to its wide variety of applications (e.g., \cite{ahn2011new,henna2013approximating,islam2009maximizing,pemmaraju2006energy}). There are two approaches to maximize the number of distinct dominating sets under the constraint on the number of times a node can appear in a dominating set -- disjoint dominating sets, and non-disjoint dominating sets.

\subsection{Disjoint Dominating Set Based Approach}
One way to approach this problem is to partition the vertex set such that each set in the partition is a dominating set, and all dominating sets are pair-wise disjoint. Such a partition is known as the \textit{domatic partition}, and the maximum number of (disjoint dominating) sets that can be obtained is known as the \textit{domatic number}, denoted by $\gamma$. Since dominating sets are pair-wise disjoint in such a partition, each vertex belongs to only one of the dominating sets. Moreover, since each node can be active for $\sigma$ time slots, each dominating set can remain active for $\sigma$ time slots. If only one dominating set is active at any time instant, which is sufficient for the complete coverage, then the lifetime of the network achievable through this approach is given by 

\begin{equation}
\label{eq:disjoint}
k = \sigma\gamma
\end{equation} 

\noindent
time slots, where $\gamma$ is the domatic number of a graph. The domatic partition problem is known to be NP-hard \cite{feige2002approximating}. Various sensor scheduling schemes that utilize domatic partitions have been proposed to maximize the network lifetime while ensuring complete coverage (e.g., \cite{pemmaraju2006energy,moscibroda2005maximizing,yu2014domatic}).

\subsection{Non-Disjoint Dominating Set Based Approach}
\begin{figure}[!htb]
\centering
\includegraphics[scale=.55]{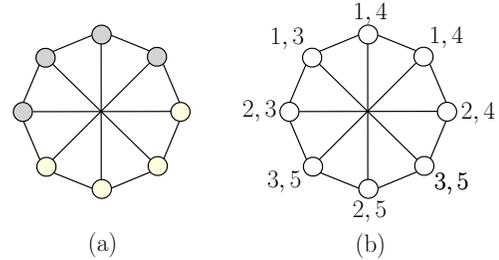}
\caption{(a) Two disjoint dominating sets are shown. (b) Five non-disjoint dominating sets, indicated by the nodes with the same labels, are shown. Each node belongs to two distinct dominating sets.}
\label{fig:nondisjoint}
\end{figure}

Another way to approach the network lifetime maximization while maintaining complete coverage is by using the \textit{non-disjoint dominating sets} of active nodes. Using this approach, it is possible to obtain a better lifetime as compared to the disjoint dominating sets based approach \cite{cardei2005energy,wang2009optimization}. As an illustration, consider the network in Fig. \ref{fig:nondisjoint}, which has a domatic number $2$. We assume that each node can be active for two time slots, i.e., $\sigma=2$, then using disjoint dominating sets approach, we get the network lifetime of $k=4$ time slots. However, it is possible to obtain five distinct dominating sets such that each node appears in at most two such sets, as shown in Fig. \ref{fig:nondisjoint}(b), thus, yielding a network lifetime of $k=5$ time slots.

The problem of finding the maximum number of dominating sets under the constraints on the number of times a node can be included in a dominating set is related to the notion of $(k,\sigma)$-configurations \cite{abbas2015deploying,fujita2000study} as defined below.

\begin{definition} \textit{($(k,\sigma)$-Configurations in Graphs)}
Let $\sigma$, $k$ be two positive integers, and $\mathcal{K}=\{1,\cdots,k\}$ be the set of labels, then $(k,\sigma)$-configuration of a graph is the assignment of $\sigma$ distinct labels from the set $\mathcal{K}$ to each node in the graph such that for every $i\in \mathcal{K}$ and every node in $v$, the label $i$ is assigned to $v$ or one of its neighbors.
\end{definition}

An example of $(5,2)$-configuration is shown in Fig. \ref{fig:nondisjoint}(b). Note that the set of nodes corresponding to a particular label in $\mathcal{K}$ constitute a dominating set. So, if a graph has an $(k,\sigma)$-configuration, it is possible to have $k$ distinct (possibly non-disjoint) dominating sets such that each node can be included in at most $\sigma$ such dominating sets. Thus, for a given $\sigma$, the maximum value of $k$, say $k^\ast$, for which $(k^\ast,\sigma)$-configuration exists, is of particular interest as it provides a scheduling scheme based on the non-disjoint dominating sets to maximize network lifetime while ensuring complete coverage.

Obviously, for $\sigma=1$, the maximum $k$ for which $(k,1)$-configuration of a graph $G$ exists, is equal to the domatic number $\gamma$ of $G$. Thus, given a $(\gamma,1)$-configuration of $G$ with the labeling set $\{\ell_1,\ell_2,\cdots,\ell_\gamma\}$, a $(k,\sigma)$-configuration could be obtained for some $\sigma>1$ and $k=\sigma\gamma$ by simply replacing each label $\ell_i$ by a set of labels $\{\ell_{i,1},\cdots,\ell_{i,\sigma}\}$. Thus, for a given $\sigma$, if $k^\ast$ is the maximum value for which $(k^\ast,\sigma)$-configuration of a graph exists, then
\begin{equation}
\label{eq:DvsND}
k^\ast \ge \sigma\gamma.
\end{equation}

Consequently, the non-disjoint dominating sets approach is always at least as good as disjoint dominating sets approach, though it often performs better. An interesting question here is under what conditions or specific instances $k^\ast>\sigma\gamma$? In this regard, first we note that every connected graph has $\gamma \ge 2$, and therefore, for a given $\sigma$, $k^{\ast}$ is always at least $2\sigma$. However, there exists many graphs for which $\gamma=2$, but $k^{\ast} > 2\sigma$. For instance, many \textit{cubic graphs}\footnote{graphs in which each vertex has a degree three.} have a domatic number of 2, e.g., the one shown in Figure \ref{fig:nondisjoint}. However, the following theorem asserts that all cubic graph have $k^{\ast}\ge \frac{5}{2}\sigma$ for a given $\sigma$.

\begin{theorem}\cite{fujita2000study}
Any cubic graph has an $(k,\sigma)$-configuration with $k=\lfloor 5\sigma/2 \rfloor$, and such a configuration can be found in polynomial time.
\end{theorem}

Recently, it has been shown in \cite{abbas2015deploying} that the above result is true even for a bigger class of graphs as stated in Theorem \ref{thm:K16}. Here, $K_{1,6}$ is a star graph with one central node of degree six, and six end nodes each with a degree one ($K_{1,6} = \star$).

\begin{theorem}\cite{abbas2015deploying}
\label{thm:K16}
Let $G$ be a graph such that 
\begin{enumerate}
\item[\textendash]
$G$ has a minimum degree at least two, 
\item[\textendash]
no subgraph of $G$ is isomorphic to $K_{1,6}$, and 
\item[\textendash]
$G \ne \{\cycle4,\c4c4,\C,\cyc,\A,\B,\D,\K\}$; 
\end{enumerate}
then $G$ has an $(k,\sigma)$-configuration with $k=\lfloor \frac{5\sigma}{2}\rfloor$.
\end{theorem}

The above result is particularly useful as proximity graphs 
(e.g., random geometric graphs), which are often used to model the limited range communication in networks such as wireless sensor networks, are always $K_{1,6}$-free. As pointed out in \cite{abbas2015deploying}, a large number of graphs in this family have a domatic number of 2, thus, non-disjoint dominating set based strategy is strictly better than the disjoint dominating set based strategy in those cases for maximizing the network lifetime while ensuring complete coverage of targets.

%% file: Related_Work.tex
\section{Related Work}
\label{sec:related_work}

One of the earliest efforts to conserve battery power through scheduling sensor devices is the work of Slijepcevic and Potkonjak~\cite{slijepcevic2001power}.
In~\cite{slijepcevic2001power}, the authors consider the problem of maximizing lifetime while preserving complete coverage of an area, which they formulate as the Set K-Cover Problem.
To solve this problem, they introduce a heuristic for finding mutually exclusive sets of sensors such that each set completely covers the monitored area.
In a follow-up work, Abrams et al.\ introduce three approximation algorithms for a variation of the Set K-Cover Problem~\cite{abrams2004set}.
Later, Deshpande et al.\ study several generalizations of the Set K-Cover Problem, and develop an approximation algorithm based on a reduction to Max K-Cut~\cite{deshpande2011energy}.

Besides the Set K-Cover Problem, researchers have studied various other formulations of the scheduling problem.
Moscibroda and Wattenhofer consider disjoint dominating-set based clustering in sensor networks~\cite{moscibroda2005maximizing}.
The authors study the problem of maximizing the lifetime of a sensor network, and provide approximation algorithms for multiple variations of the problem.
%
Cardei et al.\ study schedules that consist of non-disjoint sets of sensors and continuously monitor all targets~\cite{cardei2005energy}.
They model the solution as the maximum set covers problem, and propose two heuristics based on linear programming and a greedy approach.
Koushanfar et al.\ consider the problem of scheduling sensor devices such that the values of sleeping devices can always be recovered from the measurements of active devices within a given error bound~\cite{koushanfar2006sleeping}.
The authors first introduce a polynomial-time isotonic regression for recovering the values of sleeping devices, and  building on this regression, they then formulate the scheduling problem as domatic partitioning problem, which they solve using an ILP~solver.

Our approach is most related to the work of 
Wang et al., who study the trade-off between maximizing lifetime and minimizing ``coverage breach,'' that is, minimizing the total amount of time that each target is not covered by any sensors~\cite{wang2007minimum,wang2009optimization}.
The authors propose organizing the sensors into non-disjoint sets, and introduce an algorithm based on linear programming as well as a greedy heuristic.
In a follow-up work, Rossi et al.\ propose an exact approach based on a column-generation algorithm for solving the scheduling problem, and they also derive a heuristic from their approach~\cite{rossi2012column}. However, graph-theoretic formulation proposed in this paper allows us to directly exploit the network structure to obtain optimal schedule for a given network lifetime maximizing the detection or identification of targets. Moreover, unlike previous solutions, our game theory based solution could simultaneously solve the placement as well as scheduling problems, which gives improved performance as compared to separately solving placement and scheduling.

A few research efforts have considered simultaneous placement and scheduling.
Krause et al.\ study simultaneous placement and scheduling of sensor devices for monitoring spatial phenomena, such as road traffic~\cite{krause2011simultaneous}.
The authors assume that for any set of active sensors, the ``sensing quality'' is given by a submodular function, and they aim to maximize the worst-case sensing quality.
T{\"u}rko{\u{g}}ullar{\i} et al.\ consider the problem of maximizing lifetime through sink placement, scheduling, and determining sensor-to-sink flow paths, under energy, coverage, and budget constraints~\cite{turkougullari2010efficient}.
To solve this problem, they propose a mixed-integer linear programming model as well as a heuristic, which is more scalable but lacks performance guarantees. 

A number of studies have focused on the placement of sensor nodes, without considering sleep scheduling.
Younis and Akkaya have surveyed earlier literature on node placement, including the placement of sensor nodes~\cite{younis2008strategies}.
Krause et al.\ consider the problem of deploying sensors for detecting malicious contaminations in large-scale water-distribution networks~\cite{krause2008efficient}.
Based on the submodularity of realistic objective functions, the authors design scalable placement algorithms with provable performance guarantees.
Furthermore, they show that their method can be extended to multicriteria optimization and adversarial objectives.
Hart and Murray provide a survey of sensor placement strategies for water-distribution networks~\cite{hart2010review}.


Finally, besides scheduling, researchers have also studied other similar approaches for conserving battery power.
For example, Zhao et al.\ consider selective collaboration of sensors in order to minimize communication and communication, which increases the longevity of networks of battery-powered sensors~\cite{zhao2002information}.
The authors focus on the problem of tracking, and they study optimizing the information utility of data for given costs of communication and computation.